\documentclass[superscriptaddress, showkeys, prc]{revtex4-2}
\usepackage{amsmath,amsfonts,amssymb,amsthm,mathtools}
\usepackage{txfonts}
\usepackage{icomma}
\usepackage{fdsymbol}
\usepackage[dvipsnames]{xcolor}
\usepackage{graphicx}
\usepackage{dcolumn}
\usepackage{bm}
\usepackage{array,tabularx,tabulary,booktabs}
\usepackage{longtable}
\usepackage{multirow}
\usepackage[normalem]{ulem}

\graphicspath{{images/}}
\newcolumntype{M}[1]{>{\centering\arraybackslash}m{#1}}
\begin{document}
	
	\author{S. B. Dubovichenko}
	\affiliation{Fesenkov Astrophysical Institute "NCSRT" ASA MDASI RK$,$ 050020 Almaty$,$ Kazakhstan}
	\affiliation{al-Farabi Kazakh National University$,$ 050040 Almaty$,$ Kazakhstan}
	\author{A. S. Tkachenko}
	\affiliation{Fesenkov Astrophysical Institute "NCSRT" ASA MDASI RK$,$ 050020 Almaty$,$ Kazakhstan}
	\author{R. Ya. Kezerashvili}
	\affiliation{New York City College of Technology$,$ City University of New York$,$ Brooklyn$,$ 11201 New York$,$ USA}
	\affiliation{Graduate School and University Center$,$ City University of New York$,$ 10016 New York$,$ USA}
	\author{N. A. Burkova}
	\affiliation{al-Farabi Kazakh National University$,$ 050040 Almaty$,$ Kazakhstan}
	\author{A. V. Dzhazairov-Kakhramanov}
	\affiliation{Fesenkov Astrophysical Institute "NCSRT" ASA MDASI RK$,$ 050020 Almaty$,$ Kazakhstan}
	
	\title{The $^6$Li$(p, \gamma)^7$Be reaction rate in the light
		of the new LUNA data}
	
	\date{\today }
	
	\begin{abstract}
		We present new calculations of the astrophysical $S-$factor and reaction rate for the $^{6}$Li$(p,\gamma )^{7}$Be reaction at energies of 10 keV to 5 MeV in the framework of a modified potential cluster model with forbidden states, including low lying resonances. The astrophysical $S(E)-$factor is compared with the available experimental data and calculations done within different models. The results for the $S-$factor are in good agreement with the data set (for $E<0.3$ MeV) and calculations (for $E<0.6$ MeV) of LUNA collaboration (Phys. Rev. C 102$,$ 052802, 2020). The recommended extrapolated zero value $S(0)$ turned out to be 101 eV $\cdot $ b. Using the theoretical total cross-sections$,$ the $^{6}$Li$(p,\gamma)^{7} $Be capture reaction rate is calculated at temperatures ranging from 0.01 to 10 $T_{9}$ and compared with NACRE and NACRE II. Analytical expressions for the $S-$factor and reaction rate are given, and the effect of low-lying resonances on the reaction rate is estimated.
		We suggest to update the NACRE and NACRE II databases in light of the new LUNA data and present calculations.
	\end{abstract}
	
	\maketitle

	\preprint{APS/123-QED}
	
	%\affiliation{Fesenkov Astrophysical Institute "NCSRT" ASA MDASI RK$,$ 050020 Almaty$,$ Kazakhstan}
	%\affiliation{al-Farabi Kazakh National University$,$ 050040 Almaty$,$ Kazakhstan}
	
	%\affiliation{Fesenkov Astrophysical Institute "NCSRT" ASA MDASI RK$,$ 050020 Almaty$,$ Kazakhstan}
	
	%\affiliation{New York City College of Technology$,$ City University of New York$,$ Brooklyn$,$ 11201 New York$,$ USA}
	%\affiliation{Graduate School and University Center$,$ City University of New York$,$ 10016 New York$,$ USA}
	
	%\affiliation{al-Farabi Kazakh National University$,$ 050040 Almaty$,$ Kazakhstan}
	
	%\affiliation{Fesenkov Astrophysical Institute "NCSRT" ASA MDASI RK$,$ 050020 Almaty$,$ Kazakhstan}
	
	%\thanks{A footnote to the article title}

	\section{Introduction}
	\label{sec:introduction}
	The radiative $^{6}$Li$(p,\gamma )^{7}$Be capture reaction is of\ great interest in nuclear astrophysics \cite{Barnes1982,Boyd2010}. Since 1955 the $^{6}$Li$(p,\gamma )^{7}$Be reaction
	at low energies has been studied by several experimental groups \cite {Bashkin1955,Switkowski1979,Ostojic1983,Bruss1993,Paradellis1999,He2013,Piatti2020,Prior2004,Kiss2021}. Measurements of the astrophysical $S-$factor of this reaction were limited to the energy range of 35 keV to 1.2 MeV. The astrophysical $S-$factor and reaction rate were studied in the framework of different theoretical approaches and methods \cite{Arai2002,Huang2010,Dubovichenko2010,Dubovichenko2011,Xu2013,Dong2017,Gnech2019,Kiss2021}. A detailed review of the theoretical and experimental current status is given in Ref. \cite{Piatti2020}.
	
	In 2020, new experimental data were obtained in the Laboratory for
	Underground Nuclear Astrophysics (LUNA) \cite{Piatti2020}, and it excluded the possibility of resonance mentioned in \cite{He2013}. It seems challenging to consider this reaction in the astrophysical energy range, for which experimental data are available. In particular, our interest is the re-examination of $S-$factor and has two foci: i. to consider $^{6}$Li$(p,\gamma )^{7}$Be within the framework of the modified potential cluster model with the classification of bound and scattering states according to Young's orbital diagrams \cite{Dubovichenko2015b}; ii. to describe the $S-$factor using all available experimental data and to obtain the reaction rate. While assessing the reliability of MPCM, it is reasonable to extend the energy interval up to 5 MeV to estimate the role of resonances in this energy range.
	
	Ten years ago, we studied this reaction, but limited ourselves to an energy range up to 1 MeV. Moreover, we did not take into account resonances, nor did we consider the reaction rate \cite{Dubovichenko2010,Dubovichenko2011}.
	
	In this paper, we investigate the energy dependence of the astrophysical $S-$factor of $\ ^{6}$Li$(p,\gamma )^{7}$Be reaction at energies of 10 keV to 5
	MeV. By considering several resonances, including a wide resonance at $E_{x}=9.9
	$ MeV, the reaction rate in the temperature range of 0.01 to 10 $T_{9}$ is
	calculated. We demonstrate that it is possible to correctly convey the
	available experimental data based on potentials that are consistent with the
	energies of bound states and their asymptotic constants. For the scattering
	potentials, we use the parameters consistent with the resonance spectrum of
	the final nucleus.
	
	The results obtained for the reaction rate are approximated by curves of a
	particular type to simplify their use in applied research. These results
	apply to problems in nuclear astrophysics related to light atomic nuclei and
	ultra-low energies.
	
	This work is organized as follows. Secs. \ref{sec:theoretical_framework} and %
	\ref{sec:principles_of_the_potentials_construction} present the theoretical
	framework and fundamentals of the modified potential cluster model (MPCM), constructing principles of discrete states potentials and the description of the $p^{6}$Li channel in the continuous spectrum in the MPCM.  Cross-section for the radiative capture processes
	and discussion of wave functions asymptotics are given in Sec. \ref%
	{sec:calculation_methods}. Classification of cluster states and bound states potentials are given in Sec. \ref{sec:cluster_states_classification}, and Sec. \ref {sec:chennel_in_the_continuum} contains the description of the $p^{6}$Li channel in the continuous spectrum. Sec. \ref{sec:S-factor_and_rate} is devoted to
	the astrophysical $S-$factor and the $^{6}$Li$(p,\gamma )^{7}$Be reaction
	rate. Appendices \ref{sec:appA} and \ref{sec:appB} include the description
	of the finite-difference method that we are using in present calculations
	and a table of numerical values of the $p^{6}$Li reaction rate in
	the temperature range of 0.001 $T_{9}$ to 10 $T_{9}$, respectively. We
	outline conclusions in Sec. \ref{sec:conclusion}.
	
	\section{Theoretical framework}
	\label{sec:theoretical_framework}
	
	Charged-particle induced reactions represent one of the main inputs in
	stellar evolution. There are a number of theoretical methods used for
	description of nuclear reactions at stellar energies that are based on
	fundamental principles of quantum mechanics \cite{Descouvemont2020}.
	
	Since its first application in 1963 \cite{Tombrello1963}, the potential model approach has a special place among models for description of low energy reactions. However, over the course of over a half century, this model has been significantly modified and improved. Below we present the fundamentals of the modified potential cluster model (MPCM), where the Young diagrams are used for the classification of orbital states and construction of  potentials (\cite{Dubovichenko2015b} and Refs. herein).
	
	The basic features of the MPCM approach are as follows:
	\begin{enumerate}
		\item The MPCM is a two-particle model that accounts for the internal
		characteristics of clusters: their sizes, charges, masses, quadrupole and
		magnetic momenta, which are used to calculate the reaction total cross-sections or
		other characteristics of the final nucleus.
		
		\item The classification of cluster states is performed according to Young's
		orbital diagrams, leading to the concept of forbidden states in some partial
		waves \cite{Dubovichenko2015b}. The Pauli principle is implemented via
		exclusion of the forbidden states (FSs), manifesting in proper node behavior
		of the radial wave function (WF). Forbidden states that lead to low-lying
		bound states are not physically realized due to the orthogonality of
		corresponding functions and allowed state functions.
		
		\item The Gaussian type inter-cluster interaction potentials are
		constructed, taking into account these forbidden states in certain partial
		waves. For each partial wave with specified quantum numbers, the potential is
		constructed with two parameters, assuming it depends explicitly on
		Young's orbital diagrams.
		
		\item Potentials of the bound states (BSs) are constructed based on asymptotic constants (AC) and binding energies. Potentials of the scattering processes are constructed based on the spectra of the final nucleus or the scattering phase shifts of the particles of the input channel. Parameters of the potentials are fixed or variable within the AC error intervals and vary within the energy or width errors of resonant or excited states.
		
		\item The radial WFs of the allowed states of the continuous and discrete
		spectra are tailored appropriately using correct asymptotics.
	\end{enumerate}
	
	%It is worth noting that the correlation between the potential cluster model, the translational invariant shell model, and the nucleon association model is given in \cite{Kukulin1983}.
	
	%%% Do we need this sentence? %%%
	In light of the new experimental LUNA data \cite{Piatti2020}, we reexamine the $^6$Li$(p, \gamma)^7$Be reaction $S-$factor and reaction rate within the framework of the MPCM.
	
	\subsection*{Classification of cluster states}
	\label{Cluster_States}
	
	The total wave functions (WF) have the form of an antisymmetrized product of
	completely antisymmetric internal wave functions of clusters $\Psi (1, \dots, A_{1})=\Psi (\mathbf{R_{1}})$ and $\Psi (A_{1}+1, \dots, A)=\Psi (\mathbf{\
		R_{2}})$, multiplied by the corresponding wave function $\Phi (\mathbf{R})$
	of relative motion \cite{Wildermuth1977,Kukulin1983,Neudatchin1992}
	\begin{equation}
		\Psi =\hat{A}\{\Psi (\mathbf{R_{1}})\Psi (\mathbf{R_{2}})\Phi (\mathbf{R})\}.
		\label{eq:01wave_func}
	\end{equation}%
	In Eq. (\ref{eq:01wave_func}) $\hat{A}$ is the antisymmetrization operator
	that permutes nucleons from the clusters $A_{1}$ and $A_{2}$, $\mathbf{\
		R_{1}}$ and $\mathbf{R_{2}}$ are the center-of-mass radius-vectors of the
	clusters, and $\mathbf{R=R}_{1}\mathbf{-R}_{2}$ is the relative motion
	coordinate.
	
	The wave functions (\ref{eq:01wave_func}) are characterized by specific
	quantum numbers, including $JLS$ --- total momentum, orbital quantum
	momentum, and spin, respectively --- and Young's diagrams $\{f\}$, which determine the orbital part of WF permutation symmetry of the relative
	motion of the clusters.
	
	In the general case, the possible Young's orbital diagram $\{f\}_L$ of some
	nucleus $A(\{f\})$, consisting of two parts $A_1(\{f_1\}) + A_2(\{f_2\})$,
	is the direct outer product of Young's orbital diagrams $\{f\}_L = \{f_1\}_L \times \{f_2\}_L$ and is determined by Littlewood's
	theorem \cite{Kukulin1983,Neudatchin1992}. According to Elliot's theorem,
	each Young's diagram is associated with a certain orbital angular momentum
	or their combination.
	
	Spin-isospin diagrams are the direct inner product of the spin and isospin Young diagrams of a nucleus consisting of $A$ nucleons $\{f\}_{ST}=\{f\}_{S}\otimes \{f\}_{T}$. For a system with no more than eight particles such diagrams are provided in Table C of Ref. \cite{Itzykson1966}. A
	detailed procedure for defining the corresponding momenta can be
	found in the classical monograph \cite{Bohr1998}. Let us note that in Ref. \cite{Bohr1998} the definition for inner and outer products is reverses.
	
	The total Young's diagram of the nucleus is defined as the direct inner
	product of the orbital and spin-isospin diagram $\{f\}=\{f\}_{L}\otimes
	\{f\}_{ST}$. The total wave function of the system under antisymmetrization
	does not vanish identically, only if it contains an antisymmetric component $\{1^{N}\}$, where $N$ is the number of nucleons. In this case the conjugates $\{f\}_{L}$ and $\{f\}_{ST}$ are multiplied. Therefore, the diagrams $\{f\}_{L}$ conjugated to $\{f\}_{ST}$ are allowed in this channel. All other
	orbital symmetries are forbidden since they lead to zero total wave function
	of the particle system after antisymmetrization.
	
	\section{The potentials construction within the MPCM}
	\label{sec:principles_of_the_potentials_construction} Let us describe in more detail the procedure for constructing the intercluster partial potentials. Below we define the criteria and outline the sequence for finding parameters for the potentials and indicating their errors as well as ambiguities.
	
	\subsection{Discrete states}
	
	For the bound states of two clusters, the interaction potentials within the
	framework of the MPCM are constructed based on the requirement imposed to
	describe the main observable characteristics of such a nucleus. In this
	case, the potential parameters are fixed. It should be noted that this
	requirement is an idealized scenario that exists in the nucleus since it
	assumes that the ground state (GS) is a two-body single channel with
	probability closed to unity. First, we find the parameters of the bound state
	potentials. For the GS with a given number of bound allowed and
	forbidden states in the partial wave, these parameters are fixed
	unambiguously in terms of the binding energy, the radius of the nucleus, and
	the AC. When constructing the partial interaction potentials in the MPCM, it
	is assumed that interactions depend not only on the orbital angular momentum
	$L$ but also on the total spin $S$ and the total angular momentum $J$ of the
	system and also depend on Young's orbital diagrams. As in earlier work
	\cite{Dubovichenko2015b}, we use Gaussian interaction potentials, which
	depend on the quantum numbers $JLS$, and Young's diagrams $\{f\}_{L}$.
	Therefore, for different $JLS$, we have different values of the parameters
	of the partial potentials.
	
	The accuracy of determining the parameters of the BS potential is
	connected directly with the accuracy of the AC. The potential does not
	contain any other ambiguities, since according to Young's diagrams, the
	classification of states makes it possible to unambiguously fix the number
	of bound forbidden and allowed states in a given partial wave. The number of bound
	states ultimately determines the depth of the potential, while the width
	depends entirely on the value of the AC. If one fixes two parameters of the potential using two particular quantities --- the binding energy and the AC --- the error of the binding energy is
	seen to be much less than that of the AC.
	
	It should be noted that any calculations of the charge radius reflect the errors of the underlying model. In any model, the magnitude of such a radius depends on the
	integral of the model wave functions, thereby compounding sources of error.
	At the same time, the values of AC are determined from the asymptotic
	behavior of the model WFs at one point and contain significantly less error.
	The potentials of the BSs are constructed to obtain the best agreement
	with the values of the AC extracted independently from the experimental
	data. For more details, see Ref. \cite{Mukhamedzhanov1999}.
	
	\subsection{Continuum states}
	
	For the potentials of the continuous spectrum, the intercluster potential of
	the nonresonant scattering process for a given number of allowed and forbidden BSs in the considered partial wave is also constructed quite unambiguously based on the scattering phase shifts. The accuracy of the potential parameters sometimes as high as 20--30\%, is associated with the precision of the extracted scattering phase shifts from experimental data. For the $^{6}$Li$(p,\gamma )^{7}$Be reaction, the potential is unambiguous since the classification, according to Young's diagrams, makes it possible to fix the number of bound states. This completely determines the potential depth, and its width is determined by the shape of the scattering phase shifts.
	
	When constructing the nonresonant scattering potential based on the data for the nuclear spectra, it is difficult to estimate the accuracy of the
	parameters even for a given number of BSs. However, one can expect that it
	will not exceed the error discussed above. This potential should lead to a
	scattering phase shift close to zero or rise to a smoothly decreasing phase shift at low energies, since there are no
	resonance levels in the spectra of the nucleus.
	
	In resonance scattering, when a relatively narrow resonance is present in the partial wave at low energies for a given number of BSs, the potential is constructed completely unambiguously. The accuracy of determining the parameters of the interaction potentials is determined by the following factors. The depth of the potential depends on the resonance energy $E_x$ and the number of BSs. The width is determined by the accuracy of the experimental values of the level width $\Gamma$.
	
	The error of the parameters, approximately 5--10\%, usually does not exceed the error of the energy level width. This also applies to the construction of the partial potential from the resonant scattering phase shifts and the determination of its parameters from the spectral resonance of the nucleus \cite{Dubovichenko2015a,Dubovichenko2015b}.
	
	\section{Cross-section and WFs asymptotics}
	\label{sec:calculation_methods}
	To calculate the total cross-sections of radiative capture processes, we use the well-known formula  for the transitions of $NJ$ multipolarity \cite{Dubovichenko2015a,Dubovichenko2015b}
	\begin{equation}
		\begin{gathered} \sigma (NJ,J_f)=\frac{8\pi Ke^2}{\hbar^2\ k^3}
			\frac{\mu}{(2S_1+1)(2S_2+1)} \frac{J+1}{J[(2J+1)!!]^2}A_J^2 (NJ,K) \times \\
			\times \sum \limits_{L_i,J_i} P_J^2 (NJ,J_f,J_i)I_J^2 (k,J_f,J_i)
		\end{gathered}  \label{eq:02total_cross_sec_1}
	\end{equation}
	where the matrix elements of orbital $EJ(L)$-transitions have the
	following form $(S=S_{i}=S_{f})$
	\begin{equation}
		\label{eq:03a_matrix_elements_orbital}
		P_J^2(EJ,J_f,J_i)=\delta_{S_i S_f}[(2J+1)(2L_i+1)(2J_i+1) (2J_f+1)](L_i
		0J0|L_f 0)^2\begin{Bmatrix} L_i & S & J_i \\ J_f & J & L_f \end{Bmatrix},
	\end{equation}
	\begin{equation}
		\label{eq:03b_matrix_elements_orbital}
		A_J(EJ,K)=K^J\mu^J\left(\frac{Z_1}{m_1^J}+(-1)^J \frac{Z_2}{m_2^J}\right), \
		\ \ \  I_J(k,J_f,J_i)=\langle\chi_f|r^J|\chi_i\rangle,
	\end{equation}
	and the matrix elements of the magnetic $M1(S)$-transition are written as follows $(S=S_{i}=S_{f},L=L_{i}=L_{f})$
	\begin{equation}
		\label{eq:04a_matrix_elements_magnetic}
		P_1^2(M1,J_f,J_i)=\delta_{S_i S_f}\delta_{L_i L_f}
		[S(S+1)(2S+1) (2J_i+1)(2J_f+1)] \begin{Bmatrix} S & L & J_i \\ J_f & 1 & S
		\end{Bmatrix},
	\end{equation}
	\begin{equation}
		A_1(M1,K)= \frac{\hbar K}{m_0 c} \sqrt{3}
		\left(\muup_1\frac{m_2}{m_1+m_2}-\muup_2\frac{m_1}{m_1+m_2}\right), \ \ \ \
		I_J(k,J_f,J_i)=\langle\chi_f|r^{J-1}|\chi_i\rangle,\ J=1.
		\label{eq:04b_matrix_elements_magnetic}
	\end{equation}
	In Eqs. (\ref{eq:02total_cross_sec_1})--(\ref{eq:04b_matrix_elements_magnetic}) $K=E_{\gamma } / \hbar c$ is the wave number of the
	emitted photon with energy $E_{\gamma }$, $m_{1}$, $m_{2}$ and $\muup%
	_{1}$, $\muup_{2}$ are the masses and magnetic momenta of the clusters,
	respectively, and $\mu $ is the reduced mass of the system. Namely, in the
	present calculations for the reaction $^{6}$Li$(p,\gamma )^{7}$Be, $m_{1}\equiv m_{p}=1.00727646677$ amu, $m_{2}\equiv m_{^{6}\text{Li}}=6.0151232$ amu \cite{PML} and $\muup _{1}\equiv \muup_{p}=2.792847\muup_{0},$ $\muup_{2}\equiv \muup_{^{6}\text{Li}}=0.822\muup_{0}$ \cite{Tilley2002,Varlamov2015}, where $\muup_{0}$ is nuclear magneton, $\hbar ^{2}/m_{0}=41.4686$ MeV$\cdot $fm$^{2}$, where $m_{0} = 931.494$ MeV is the atomic mass unit (amu).
	
	The point-like Coulomb potential is of the form $V_{\text{Coul}}\text{(MeV)}=1.439975Z_{1}Z_{2}/r$, where $r$ is the relative distance between the particles of the channel in fm, and $Z_{1}$ and $Z_{2}$ are the charges in units of the elementary charge. The Coulomb parameter $\eta =\mu Z_{1}Z_{2}e^{2}/k\hbar ^{2}$ is represented in the form $\eta = 3.44476\cdot10^{-2}\ Z_{1}Z_{2}\mu/k$, where $k$ is the wavenumber in fm$^{-1}$
	and is determined by the energy $E_{c.m.}$ of the interacting particles, $k^{2}=2\mu E_{c.m.}/\hbar ^{2}$.
	
	In our calculations, we use the dimensionless AC denoted as $C_{\text{w}}$
	\cite{Plattner1981}
	\begin{equation}
		\chi _{L}(r)=\sqrt{2k_{0}}\ C_{\text{w}}\ W_{-\eta L+1/2}(2k_{0}r).
		\label{eq:05dimensionless_AC}
	\end{equation}%
	A dimensional asymptotic constant $C$ is related to the
	asymptotic normalization coefficient (ANC) $A_{NC}$ by the expression \cite%
	{Mukhamedzhanov1999}
	\begin{equation}
		A_{NC}=\sqrt{S_{F}}C,
		\label{eq:06AC_ANC}
	\end{equation}%
	where $S_{F}$ is the spectroscopic factor and $C$ is the dimensional AC that
	can be represented using the asymptotics of the WF
	\begin{equation}
		\chi _{L}(r)\xrightarrow[r \rightarrow R]{}CW_{-\eta L+1/2}\left(
		2k_{0}r\right) .
		\label{eq:07}
	\end{equation}%
	In Eq. (\ref{eq:07}) $R$ is the large distance where the nuclear potential vanishes and $\chi _{L}(r)$ is the wave function of the bound state obtained from the solution of the radial Schr$\ddot{\text{o}}$dinger equation and normalized to unity. The Whittaker function $W_{-\eta L+1/2}$ of the bound
	state determines the asymptotic behaviour of the WF. The wave number $k_{0}$
	is related to the channel binding energy $E_{b}$.
	
	For a continuous spectrum, the function $\chi_i$ found numerically is
	matched to asymptotics $u_L(R)$ of the form
	\begin{equation}  \label{eq:08}
		N_L u_L(r) \xrightarrow[r \rightarrow R]{} F_L(kr) + \tan (\deltaup%
		_{S,L}^J)G_L(kr).
	\end{equation}
	Here $F_L$ and $G_L$ are Coulomb regular and irregular functions \cite%
	{Abramowitz1972}. They are the solutions of the Schr$\ddot{\text o}$dinger
	equation with the Coulomb potential. $\deltaup_{S,L}^J$ are the
	scattering phase shifts depending on the $JLS$ momenta of the system and $%
	N_L $ is the normalizing constant of the numerical radial function $u_L(R)$
	for the continuum.
	
	\section{Cluster states classification and the BS potentials}
	
	\label{sec:cluster_states_classification} Consider the classification of the
	BSs of the $p^{6}$Li system according to Young's diagrams. In this case
	there is only one Young's orbital diagram for the $^{6}$Li nucleus $\{42\}$.
	It is believed that the system's potentials are dependent on the diagrams or
	combinations of these diagrams in various states. Thus, if the orbital
	diagram $\{42\}$ allowed in the $^{2}$H$^{4}$He-cluster channel is accepted
	for the $^{6}$Li nucleus, then the $p^{6}$Li system with spin $S=1/2$
	contains a forbidden level with diagram $\{52\}$ and orbital momenta of $%
	L=0, 2$, and the allowed states with configurations $\{43\}$ for $L=1, 3$ and $%
	\{421\}$ for $L=1, 2$. Hence, the $p^{6}$Li potentials must have a forbidden
	state related to $\{52\}$ in the $S$ wave. The allowed bound state corresponds
	to the $P$ wave with the two Young's diagrams $\{43\}$ and $\{421\}$. In the
	quartet spin channel $S=3/2$ of the system, only one diagram $\{421\}$ is
	allowed for $L=1, 2$ \cite{Dubovichenko2015b}. Since there are two allowed diagrams $\{43\}$ and $\{421\}$ in the doublet spin state of the $p^{6}$Li system, the scattering states turn out to be mixed in orbital symmetries. At the same time, only one allowed diagram
	$\{43\}$ usually corresponds to the doublet ground state of the $^7$Be
	nucleus in the $p^{6}$Li channel with $J^{\pi}=3/2^-$ and $L=1$.
	
	Here, the $p^{6}$Li system is completely analogous to the $p^{2}$H channel
	in the $^{3}$He nucleus. In the latter case the doublet state is also mixed
	according to Young's diagrams $\{3\}$ and $\{21\}$ \cite{Neudatchin1992}.
	Therefore, the potentials constructed based on the elastic scattering phase
	shifts of the $p^{6}$Li system or the level spectra of the $^{7}$Be nucleus
	cannot be used to describe the GS of the $^{7}$Be nucleus in the $p^{6}$Li channel. Pure in orbital
	symmetry with Young's diagram $\{43\}$, the $^{2}P_{3/2}$ potential of the
	ground state of $^{7}$Be reproduces the binding energy of the GS of the
	nucleus consistent with the $p^{6}$Li system and its asymptotic constant.
	
	The scattering potentials are constructed based on the spectra of the $^7$Be
	nucleus from Ref. \cite{Tilley2002}; it has no major difference from the
	newer compilation \cite{Sukhoruchkin2016}.
	
	The orbital state's classification of the $p^{6}$Li system is shown  in Table \ref{tab:orb_states_classification}.
	%It differs from the previous one \cite{Dubovichenko2010,Dubovichenko2011}, and now the $S$ wave has only one FS with $\{52\}$, whereas the $P$ wave has no FSs.
	
	\setlength{\extrarowheight}{2pt} %Extra space for table rows
	%\begin{widetext}
	\begin{table*}[ht]
		\caption[The classification of the orbital states of the $p^{6}$Li system
		\cite{Dubovichenko2010,Dubovichenko2015a}]{The classification of the
			orbital states of the $p^{6}$Li system \cite
			{Dubovichenko2010,Dubovichenko2015a}. The following notations are used: $S$
			and $L$ are spin and orbital angular momentum of the system, respectively, $\{f\}_{S}$, $%
			\{f\}_{T}$ and $\{f\}_{ST}$ for isospin $T=1/2$, and $\{f\}_{L}$ are spin,
			isospin and spin-isospin, and possible orbital Young's diagrams, and $\{f\}_{%
				\text{AS}}$, $\{f\}_{\text{FS}}$ are Young's diagrams of allowed and
			forbidden orbital states.}
		\label{tab:orb_states_classification}
		\begin{ruledtabular}
			\begin{tabular}{cccccccc}
				$S$ & $\{f\}_S$ & $\{f\}_T$ & $\{f\}_{ST}$ & $\{f\}_L$ & $L$ & $\{f\}_{\text{AS}}$ & $\{f\}_{\text{FS}}$ \\
				\hline
				\multirow{3}{*}{$1/2$} & \multirow{3}{*}{\{43\}} & \multirow{3}{*}{\{43\}} & \{7\}+\{61\}+\{52\}+\{511\}+\{43\}+ & \{52\} & $0, 2$ &---& \{52\} \\
				&  &  & +\{421\}+\{331\}+\{4111\}+ & \{43\} & $1, 3$ &\{43\}& ---\\
				&  &  & +\{322\}+\{3211\}+\{2221\} & \{421\} & $1, 2$ &\{421\} &---\\
				\hline
				\multirow{3}{*}{$3/2$} & \multirow{3}{*}{\{52\}} & \multirow{3}{*}{\{43\}} & \{61\}+\{52\}+\{511\}+ & \{52\} & $0, 2$ &  --- & \{52\} \\
				&  &  & +\{43\}+2\{431\}+\{331\}+ & \{43\} & $1, 3$ & --- & \{43\} \\
				&  &  & +\{322\}+\{3211\} & \{421\} & $1, 2$ & \{421\} & --- \\
			\end{tabular}
		\end{ruledtabular}
		%\normalsize
	\end{table*}
	%\end{widetext}
	
	We use a Gaussian potential that depends on the momenta of the system \cite%
	{Dubovichenko2015b} and Young's diagrams
	\begin{equation}
		V(r,JLS,\{f_{L}\})=-V_{0}(JLS,\{f_{L}\})\exp (-\alpha _{JLS,\{f_{L}\}}r^{2}).
		\label{eq:09Gaussian_potential}
	\end{equation}%
	In Eq. (\ref{eq:09Gaussian_potential}) $V_{0}$ is the potential depth and $\alpha $ is related to the potential width. The choice of parameters for the
	bound and scattering states is discussed in detail in Sec. \ref{sec:chennel_in_the_continuum}.
	
	A compilation of the AC data for the ground and first excited state in the $p^{6}$Li channel of $^{7}$Be is presented in Table \ref{tab:ANC_values}, with $C_{\text{w}}^{2}=A_{NC}^{2}/2k_{0}S_{F}$. Here $\sqrt{2k_{0}}=0.983$ fm$^{-1/2}$ for the GS and $\sqrt{2k_{0}}=0.963$ fm$^{-1/2}$ for the FES.
	\begin{widetext}
		\begin{table*}[h]
			\caption[AC data for the ground and the first excited states in the $p^{6}$Li channel of $^7$Be.]{AC data for the ground and the first excited states in $p^{6}$Li channel of $^7$Be.}
			\label{tab:ANC_values}
			\begin{ruledtabular}
				\begin{tabular}{M{0.1\textwidth}m{0.3\textwidth}M{0.2\textwidth}M{0.2\textwidth}M{0.13\textwidth}}
					BS & \hfil Reference & $A_{NC}$, fm$^{-1/2}$ & $S_F$ & $C_{\text w}$ \\
					\hline
					\multirow{6}{2em}{GS} & Nollett and Wiringa \cite{Nollett2011}, 2011 & 2.85(3) & 1 & 2.90(3) \\
					& Huang \textit{et al.} \cite{Huang2010}, 2010 & 2.01 & 0.66 -- 1 & 2.28(24) \\
					& Timofeyuk \cite{Timofeyuk2013}, 2013 & 1.80 & 0.46 -- 0.87 & \textbf{2.32(37)} \\
					& Burtebayev \textit{et al.} \cite{Burtebayev2013}, 2013 & 1.77(8) & 0.55 -- 0.81 & \textbf{2.23(31)} \\
					& Gnech and Marcucci \cite{Gnech2019}, 2019 & 2.654 & 1.003 & 2.65 \\
					& Kiss\textit{ et al.} \cite{Kiss2021}, 2021 & 2.19(9) & 0.98(30) & 2.35(43) \\
					\hline
					\multirow{5}{2em}{FES} & Huang \textit{et al.} \cite{Huang2010}, 2010 & 1.91 & 0.66 -- 1.02 & 2.20(24) \\
					& Timofeyuk \cite{Timofeyuk2013}, 2013 & 1.91 & 0.62 -- 1.21 & \textbf{2.17(36)} \\
					& Burtebayev \textit{et al.} \cite{Burtebayev2013}, 2013 & 1.95(9) & 0.85 -- 1.03 & \textbf{2.10(20)} \\
					& Gnech and Marcucci \cite{Gnech2019}, 2019 & 2.528 & 1.131 & 2.53 \\
					& Kiss \textit{et al.} \cite{Kiss2021}, 2021 & 2.18(6) & 1.08(32) & 2.26(39) \\
				\end{tabular}
			\end{ruledtabular}
		\end{table*}
	\end{widetext}
	
	Summarizing the data in Table \ref{tab:ANC_values}, we conclude that
	all cited data on the AC are overlapped. While constructing the corresponding GS
	and FES potentials, we used the average values indicated in bold font in Table \ref{tab:ANC_values}. Just at the end of our calculation story, a
	publication by Kiss \textit{et al.} \cite{Kiss2021} appeared, and it happened that these
	latest experimental results turned to be within the defined intervals
	for ANC given above.
	
	We use GS and FES in the form of only doublet $^{2}P_{3/2}$ and $^{2}P_{1/2}$
	states, but we take the experimental data on ANC from \cite{Burtebayev2013}
	since it is assumed that these states result in the observed ANC values. We
	do not consider these states as a mix of doublet and quartet states, for
	instance, $^{2+4}P_{3/2}$, is a prime example, as the quartet channel is not
	allowed for the orbital Young's diagram  $\{43\}$ in Table \ref{tab:orb_states_classification}.
	
	All AC values are used here as a framework to obtain the parameters of
	the $p^{6}$Li interaction BSs potentials. These potentials correspond to the
	lower, upper, and average values of AC and accurately reproduce the binding
	energies \cite{Tilley2002} of the bound states. The parameters of the
	potentials are presented in Table \ref{tab:Parameters}.
	
	\begin{widetext}
		\begin{table*}[h]
			\caption[Parameters of the $p^{6}$Li system bound state potentials and characteristics]{Parameters of $p^{6}$Li system bound state potentials and bound states characteristics. $E_x$, $E_b$, and $V_0$ are provided in MeV. The $R_{\text{ch}}$ and $R_{\text{m}}$ are given in fm.}
			\label{tab:Parameters}
			\begin{ruledtabular}
				\begin{tabular}{M{0.05\textwidth} M{0.05\textwidth} M{0.1\textwidth} M{0.07\textwidth} M{0.1\textwidth} M{0.07\textwidth} M{0.1\textwidth} M{0.1\textwidth} M{0.1\textwidth} M{0.1\textwidth} M{0.05\textwidth}}
					$\#$ & BS & $E_x$ & $J^{\pi}$ & $E_b$ & $^{2S+1}L_J$ & $V_0$ & $\alpha$, fm$^{-2}$ & $C_\text{w}$ & $R_{\text{ch}}$ &$R_{\text{m}}$ \\
					\hline
					1& GS & 0 & 3/2$^-$ & –5.60580 & $^2 P_{3/2}$ & $100.750920$ & $0.25$ & $1.75(1)$ & $2.49$ & $2.51$ \\
					
					2& GS & 0 & 3/2$^-$ & –5.60580 & $^2 P_{3/2}$ & $74.504070$ & $0.17$ & $2.26(1)$ & $2.58$ & $2.58$ \\
					
					3& GS & 0 & 3/2$^-$ & –5.60580 & $^2 P_{3/2}$ & $60.998575$ & $0.13$ & $2.74(1)$ & $2.64$ & $2.61$ \\
					
					4& FES & 0.4291 & 1/2$^-$ & –5.17670 & $^2 P_{1/2}$ & $99.473500$ & $0.25$ & $1.68(1)$ & $2.52$ & $2.54$ \\
					
					5& FES & 0.4291 & 1/2$^-$ & –5.17670 & $^2 P_{1/2}$ & $73.333835$ & $0.17$ & $2.16(1)$ & $2.59$ & $2.59$ \\
					
					6& FES & 0.4291 & 1/2$^-$ & –5.17670 & $^2 P_{1/2}$ & $59.898120$ & $0.13$ & $2.61(1)$ & $2.65$ & $2.62$ \\
				\end{tabular}
			\end{ruledtabular}
		\end{table*}
	\end{widetext}
	
	%These potentials accurately reproduce the binding energies \cite{Tilley2002} of the bound states.
	
	\section{$\boldmath p^6 \text{Li}$ channel in the continuous spectrum}
	
	\label{sec:chennel_in_the_continuum} We usually assume that the scattering
	potentials can lead to the FSs \cite{Dubovichenko2015b}, and if the FSs are
	absent, the potential depth can be set to zero. The present case refers to
	the scattering $P$ potentials without the FS, while the $S$ and $D$
	potentials have the bound forbidden state and, even at zero phase shifts,
	must have a nonzero depth.
	
\begin{figure}[h!]
		\includegraphics[width=0.5\textwidth]{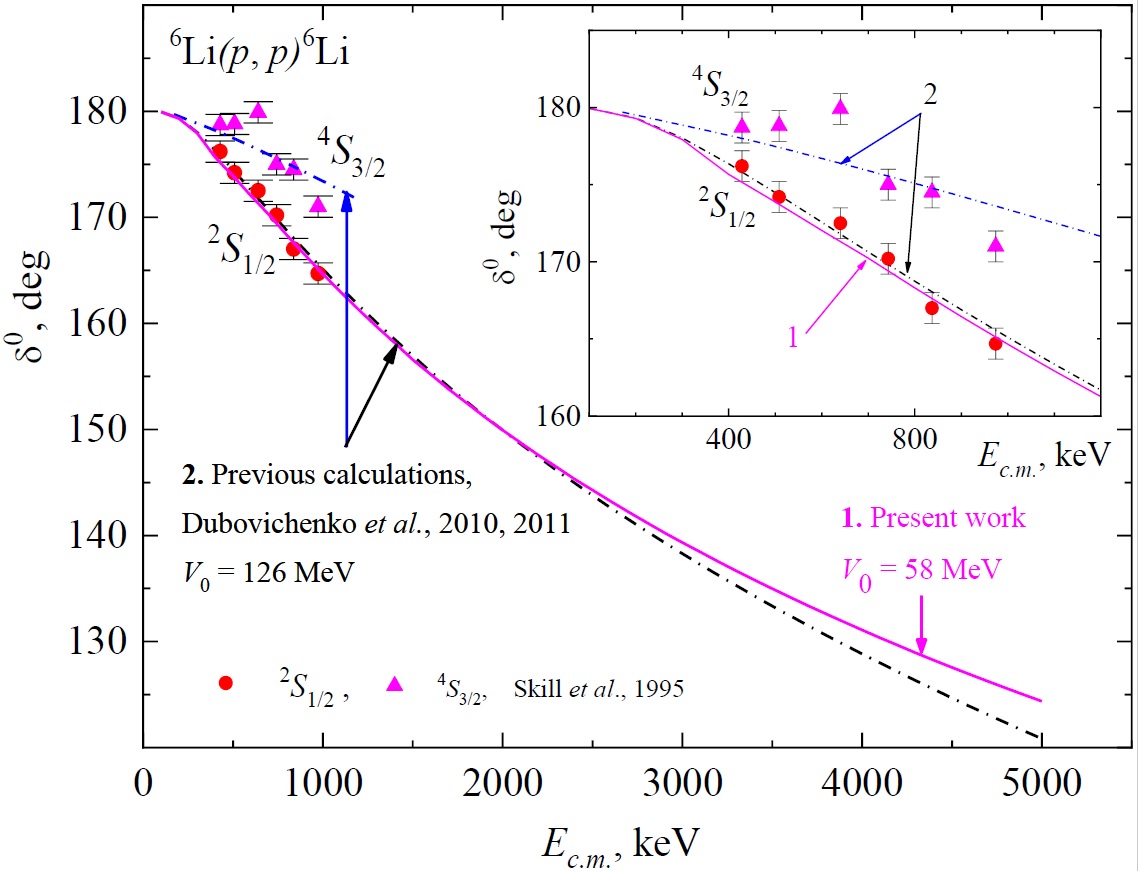}
		\caption{Doublet and quartet $S$ phase shifts of elastic $p^{6}$Li
			scattering at low energies. The $^{2}S$ and $^{4}S$ phase shifts are taken from Ref. \cite{Skill1995} and shown by {\color{Red}$\medblackcircle$} and {\color{Magenta}$\medblacktriangleup$}, respectively. Results from \cite{Dubovichenko2010,Dubovichenko2011} obtained according to \cite{Skill1995} are shown by the dashed-dotted curves. Results of the present work are shown by the solid curve.}
		\label{fig:phase_shifts_S}
	\end{figure}

\begin{figure}[th]
		\includegraphics[width=0.5\textwidth]{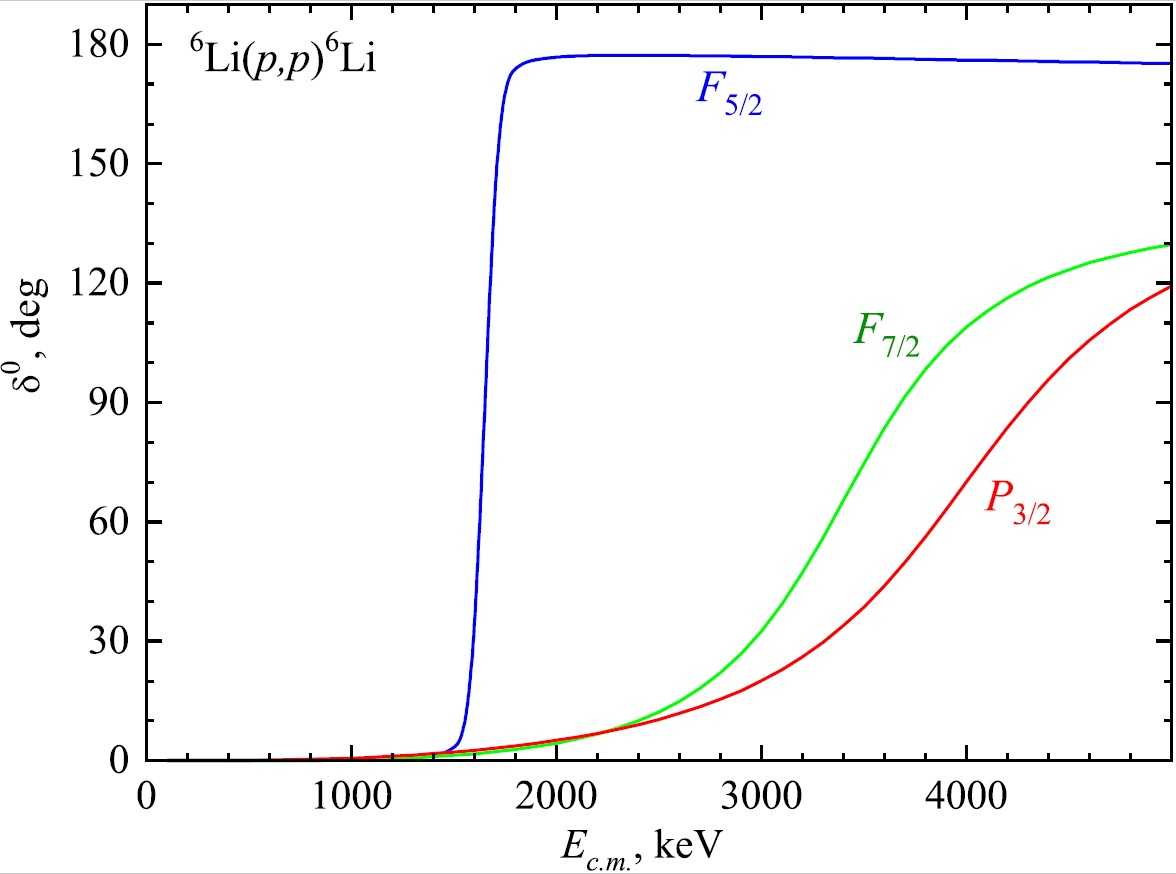}
		\caption{$P$ and $F$ phase shifts of elastic $p+^{6}$Li scattering obtained
			for scattering potentials with the parameters from Tables \ref
			{tab:spectrumGS} and \ref{tab:spectrumFES}.}
		\label{fig:phase_shifts_FP}
	\end{figure}

	In the presence of one BS, the phase shift starts at $180^{\circ}$ \cite%
	{Neudatchin1992}, as shown in Fig. \ref{fig:phase_shifts_S} for the $S$
	scattering phase shifts. Furthermore, we consider $E2$ transitions from
	resonant $F$ scattering states with phase shifts shown in Fig. %
	\ref{fig:phase_shifts_FP}. The parameters of potentials for all scattering
	processes for transitions to GS and FES are given in Tables \ref%
	{tab:spectrumGS} and \ref{tab:spectrumFES}, respectively.
	
	% The potentials for all scattering processes are
	% given in Table \ref{tab:spectrumGS} for transitions to GS and in Table \ref{tab:spectrumFES} for transitions to FES.
	
	\begin{widetext}
		\begin{table*}[h]
			\caption[The spectrum of $^7$Be energy levels and scattering states in the $p^{6}$Li channel, GS]{The spectrum of $^7$Be levels \cite{Tilley2002} and scattering states in the $p^{6}$Li channel for the capture to the $^2 P_{3/2}$ GS at a binding energy of 5.6058 MeV, along with $P^2_J$ from expressions (\ref{eq:03a_matrix_elements_orbital}) and (\ref{eq:04a_matrix_elements_magnetic}). $E_x$, $E_{\text {res}}$, $\Gamma_{c.m.}$ and $V_0$ are provided in MeV.}
			\label{tab:spectrumGS}
			\small
			\begin{ruledtabular}
				\begin{tabular}{ccccccccccc}	
					\multirow{2}{*}{\#}& $E_x$, & \multirow{2}{*}{$J^{\pi}$} & $E_{\text {res}}$, & $\Gamma_{c.m.}$, & GS transition: & \multirow{2}{*}{$P^2_J$} & \multirow{2}{*}{$V_0$} & $\alpha$, & $E_{\text {res}}$ & $\Gamma_{c.m.}$ \\
					
					& exp. &  & exp. & exp. & $\left[^{2S+1}L_J\right]_i \rightarrow \left[^{2S+1}L_J\right]_f$ &  &  & fm$^{-2}$ & theory & theory \\
					\hline
					1 & No res. & $5/2^+$ & --- & --- & $E1:\ ^2D_{5/2} \rightarrow \ ^2P_{3/2}$ & $36/5$ & $58.0$ & $0.4$ & --- & --- \\
					
					2 & No res. & $3/2^+$ & --- & --- & $E1:\ ^2D_{3/2} \rightarrow \ ^2P_{3/2}$ & $4/5$ & $58.0$ & $0.4$ & --- & --- \\
					
					3 & No res. & $1/2^+$ & --- & --- & $E1:\ ^2D_{5/2} \rightarrow \ ^2P_{3/2}$ & $4$ & $58.0$ & $0.4$ & --- & --- \\
					
					4 & No res. & $1/2^-$ & --- & --- & $M1:\ ^2P_{1/2} \rightarrow \ ^2P_{3/2}$ & $4/3$ & $0.0$ & $1.0$ & --- & --- \\
					\hline
					5 & $7.2(1)$ & $5/2^-$ & $1.59(10)$ & $0.40(5)$ & $E2:\ ^2F_{5/2} \rightarrow \ ^2P_{3/2}$ & $12/7$ & $111.60$ & $0.1$ & $1.60(1)$ & $0.62(1)$ \\
					
					6 & $9.29(31)$ & $7/2^-$ & $3.68(31)$ & $1.93(96)$ & $E2:\ ^2F_{7/2} \rightarrow \ ^2P_{3/2}$ & $72/7$ & $44.34$ & $0.05$ & $3.68(1)$ & $1.50(1)$ \\
					
					7 & $9.9$ & $3/2^-$ & $4.3$ & $1.8$ & $M1:\ ^2P_{3/2} \rightarrow \ ^2P_{3/2}$ & $5/3$ & $432.0$ & $1.5$ & $4.30(1)$ & $1.80(2)$ \\
					
				\end{tabular}
			\end{ruledtabular}
			\normalsize
		\end{table*}
	\end{widetext}
	
	\begin{widetext}	
		\begin{table*}[h]
			\caption[The spectrum of $^7$Be energy levels and scattering states in the $p^{6}$Li channel, FES]{The spectrum of $^7$Be energy levels \cite{Tilley2002} and scattering states in the $p^{6}$Li channel for the proton capture to the $^2 P_{1/2}$ FES at a binding energy of 5.1767 MeV, along with $P^2_J$ from expressions (\ref{eq:03a_matrix_elements_orbital}) and (\ref{eq:04a_matrix_elements_magnetic}). $E_{\text x}$, $E_{\text {res}}$, $\Gamma_{c.m.}$ and $V_0$ are provided in MeV.}
			\label{tab:spectrumFES}
			\begin{ruledtabular}
				\begin{tabular}{ccccccccccc}
					\multirow{2}{*}{\#}& $E_{\text x}$, & \multirow{2}{*}{$J^{\pi}$} & $E_{\text {res}}$, & $\Gamma_{c.m.}$, & FES transition: & \multirow{2}{*}{$P^2_J$} & \multirow{2}{*}{$V_0$} & $\alpha$, & $E_{\text {res}}$ & $\Gamma_{c.m.}$ \\
					
					& exp. &  & exp. & exp. & $\left[^{2S+1}L_J\right]_i \rightarrow \left[^{2S+1}L_J\right]_f$ &  &  & fm$^{-2}$ & theory & theory \\
					\hline
					1 & No res. & $3/2^+$ & --- & --- & $E1:\ ^2D_{3/2} \rightarrow \ ^2P_{1/2}$ & $4$ & $58.0$ & $0.4$ & --- & --- \\
					
					2 & No res. & $1/2^+$ & --- & --- & $E1:\ ^2S_{1/2} \rightarrow \ ^2P_{1/2}$ & $2$ & $58.0$ & $0.4$ & --- & --- \\
					
					3 & No res. & $1/2^-$ & --- & --- & $M1:\ ^2P_{1/2} \rightarrow \ ^2P_{1/2}$ & $1/6$ & $0.0$ & $1.0$ & --- & --- \\
					\hline
					4 & $7.2(1)$ & $5/2^-$ & $1.59(10)$ & $0.40(5)$ & $E2:\ ^2F_{5/2} \rightarrow \ ^2P_{1/2}$ & $6$ & $111.6$ & $0.1$ & $1.60(1)$ & $0.62(1)$ \\
					
					5 & $9.9$ & $3/2^-$ & $4.3$ & $1.8$ & $M1:\ ^2P_{3/2} \rightarrow \ ^2P_{1/2}$ & $4/3$ & $432.0$ & $1.5$ & $4.30(1)$ & $1.80(2)$ \\
				\end{tabular}
			\end{ruledtabular}
			\normalsize
		\end{table*}
	\end{widetext}
	
\begin{figure}[th]
		\includegraphics[width=0.6\textwidth]{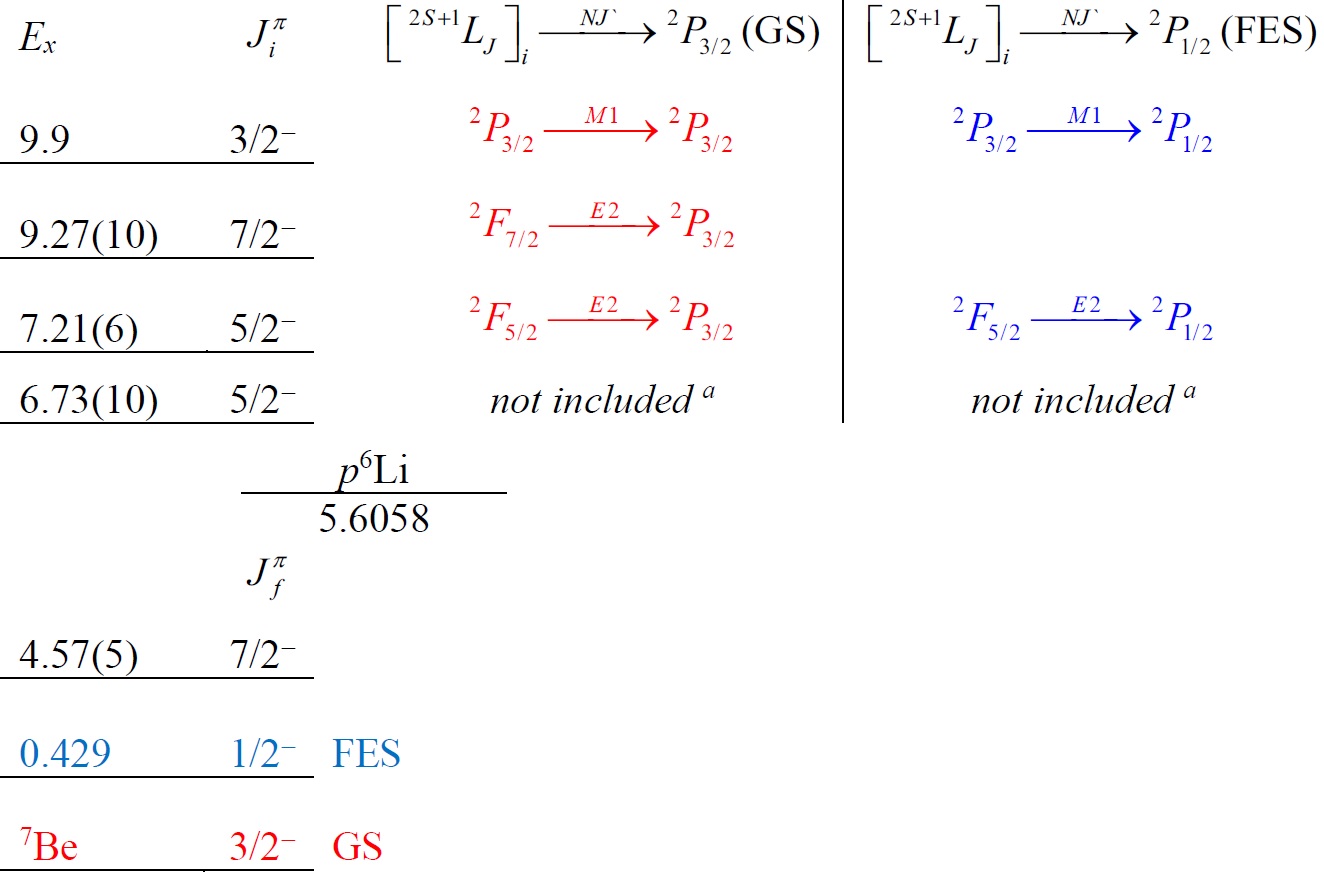}
		\caption{Schematics of the energy spectrum of $^{7}$Be. The energy are given
			in MeV and figure is not drown to scale \cite{Tilley2002}. $^{a}$
			the above-threshold resonance at 6.73 MeV with a width of 1.2 MeV refers to
			the $^{4}$He$^{3}$He channel that is not considered in the present work.}
		\label{fig:energy_spectrum}
	\end{figure}

	In addition, we consider the resonance at an excitation energy of 9.9 MeV
	\cite{Tilley2002} according to Fig. \ref{fig:energy_spectrum} (4.3 MeV above
	the threshold) in a $^{2}P_{3/2}$ scattering state of width 1.8 MeV in c.m.
	Considering such an $M1$ transition to the $^{2}P_{3/2}$ ground state or an $M1$ transition from the $^{2}P_{1/2}$ scattering state to the $^{2}P_{1/2}$ FES are possible due to the presence of different Young's diagrams in the bound and scattering states. Recall that the BSs have the diagram $\{43\}$, and the scattering states are mixed according to the two diagrams $\{43\}+\{421\}$ \cite{Dubovichenko2011}.
	
	Table \ref{tab:spectrumGS} shows possible transitions to the $^{7}$Be nucleus
	GS from various $p^{6}$Li scattering states with $^{2S+1}L_{J}$. The
	possible transitions to the FES from different scattering states with $^{2S+1}L_{J}$
	are shown in Table \ref{tab:spectrumFES}. The resonance energies and widths are
	obtained with the corresponding parameters of the scattering potentials. For
	the $P_{1/2}$ scattering wave, zero-depth potentials are used since the
	scattering $P$ waves do not contain forbidden BSs. For the $^{2}D$ wave
	potentials, $^{2}S$ wave parameters are used for $L=2$.
	
	Resonant phase shifts of elastic scattering $p+^{6}$Li
	%from Tables \ref{tab:spectrumGS} and \ref{tab:spectrumFES}
	are shown in Fig. \ref{fig:phase_shifts_FP}. The above-threshold resonance
	at 6.73 MeV with a width of 1.2 MeV indicated in Fig. \ref%
	{fig:energy_spectrum} refers to the $^{4}$He$^{3}$He channel \cite%
	{Tilley2002} and is not considered in our previous works \cite{Dubovichenko2010,Dubovichenko2011}.
	Note again that in the MPCM we used, Young's orbital diagram \{43\} is forbidden in the quartet state, as shown in Table \ref{tab:orb_states_classification}, and this particular diagram corresponds to the GS of the $^7$Li nucleus. Therefore, in the GS there is only a doublet $^2P_{3/2}$ state (without impurity of $^4P_{3/2}$), which is allowed for the diagram \{43\}. Thus, our model \cite{Dubovichenko2010,Dubovichenko2011} predicted the absence of resonance at 6.73 MeV with $J^{\pi} = 5/2$ in the nucleon channel, or, in other words, the impossibility of the $M1$ transition from this resonance to the GS. This has been confirmed by the new LUNA results \cite{Piatti2020} and, indirectly, by the data of \cite{Tilley2002}.
	The width of the resonance peak at 9.29 MeV is taken from Table 7.10 of \cite{Tilley2002},
	although another state, $^{2}P_{1/2}$, is indicated therein. At
	9.27 MeV, the given moment is $7/2$, as per Table 7.7 in Ref. \cite{Tilley2002}, so we infer the presence of an $F$ state. However, this
	resonance leads to a minimal increase in cross-sections at the $E2$
	transition. It is negligible against the background of the $E1$ resonance at
	4.3 MeV with a transition from the $^{2}P_{3/2}$ scattering state for the potential parameters \#7 (Table \ref{tab:spectrumGS}) or \#5 (Table \ref{tab:spectrumFES}), respectively.
	
	In Fig. \ref{fig:phase_shifts_S} are shown the doublet and quartet $S$ phase shifts of elastic $p^6$Li scattering at low energies. The $^{2}S$ potential from Table \ref{tab:spectrumGS} with a depth of 58 MeV
	has one FS and allows one to describe the phase shifts of \cite{Dubovichenko2010} up to 1 MeV, shown in Fig. \ref{fig:phase_shifts_S} by
	solid circles. Moreover, it gives phase shifts below 2 MeV that coincide with the phase
	shifts obtained with the potential from Ref. \cite{Dubovichenko2010}, with a depth of
	126 MeV and a width of 0.15 fm$^{-2}$. This early potential has two FSs and
	does not agree with our new classification from Table \ref{tab:orb_states_classification}. To compare the results, we construct a new potential (depth 58 MeV) that gives the most overlap in phase shifts from prior work \cite{Dubovichenko2010}. The phase shift of the new potential is given in Fig. \ref{fig:phase_shifts_S} by a solid curve, while dash-dotted curves refers to
	results from Ref. \cite{Dubovichenko2010}.
	
	\section{Astrophysical \textit{S}-factor and reaction rate}
	
	\label{sec:S-factor_and_rate}
	
	A special feature of cross-sections of nuclear reactions with charged
	particles at low and ultra-low energies is an extreme reduction by several
	orders of a cross-section magnitude due to the decrease in transmission probability
	through the Coulomb barrier. For practical purposes, the astrophysical
	S-factor is introduced as
	\begin{equation}
		S(E)=\dfrac{\sigma (E)}{P(E)}E,\ \ \ P(E)=e^{-2\pi \eta },
		\label{eq:10S-factor}
	\end{equation}%
	where the factor $P(E)$ reflects the permeability of the Coulomb barrier.
	The advantage of the astrophysical $S-$factor is that it shows a smooth
	energy dependence at low energies.
	
	Following the excellent manuscript by Christian Iliadis \cite{Iliadis2015},
	we would like to provide a brief discussion of the use of the $S-$factor in
	conventional calculation schemes in order to clarify the current approach.
	The definition of the $S-$factor (\ref{eq:10S-factor}) allows to write the
	following expression for the reaction rate:
	\begin{equation}
		N_{A}\langle \sigma \nu \rangle =\left( \dfrac{8}{\pi \mu }\right)
		^{1/2}N_{A}\left( \kappa T_{9}\right) ^{-3/2}\int\limits_{0}^{\infty
		}e^{-2\pi \eta }S(E)e^{-E/\kappa T_{9}}dE.
		\label{eq:11Reaction_rate}
	\end{equation}
	In Eq.(\ref{eq:11Reaction_rate}) $\kappa $ is the Boltzmann constant and $%
	N_{A}$ is Avogadro's number. A notable effort has been expended to bring the
	integral in (\ref{eq:11Reaction_rate}) to analytical form. This is possible
	only if the expansion of the $S(E)-$factor in the $E$ series, given by
	\begin{equation}
		S(E)\approx S(0)+S^{\prime }(0)E+S^{\prime \prime}E^2.
		\label{eq:12S-factor_expansion}
	\end{equation}%
	is valid at low energies (c.f. Ref. \cite{Iliadis2015}, Section 3.2).
	Expression (\ref{eq:12S-factor_expansion}) explains the active interest to determine the value $S(0)$ as a key one for the calculation of the
	reaction rate in the form of (\ref{eq:11Reaction_rate}) with its further analytical
	parameterizations.
	
	Table \ref{tab:S-factor_experimental} presents the available experimental data
	of the astrophysical $S-$factor for the $^{6}$Li$(p,\gamma )^{7}$Be
	reaction, as well as the extrapolated $S(0)$ values. Results of previous
	theoretical calculations and the present work are presented in Table \ref{tab:S-factor_theory}. Our calculations are analyzed and interpreted based
	on the experimental data \cite{Piatti2020} as one of the newest and most
	accurate.
	
	In the present work we introduce some corrections to \cite{Dubovichenko2010,Dubovichenko2011} that
	allow us to extend the energy interval for the cross-sections and
	corresponding $S-$factors. It is also worth mentioning that in Ref. \cite%
	{Gnech2019}, the authors used the calculation scheme based on \cite%
	{Dubovichenko2011}. The value $S(0)=95.0$ eV$\cdot $b is obtained which is
	consistent with the LUNA experimental data \cite{Piatti2020}. However, the
	astrophysical reaction rate is missing in Ref. \cite{Gnech2019}.
	
	\begin{widetext}
		\begin{table*}[h!]
			\caption[Experimental data on the astrophysical $S$-factor of $\ ^6$Li$(p, \gamma)^7$Be.]{Experimental data on the astrophysical $S-$factor of$^6$Li$(p, \gamma)^7$Be. $S(E)$ and $S(0)$ are given in eV$\cdot$b and $S(0)$ are extrapolated values.}
			\label{tab:S-factor_experimental}
			\begin{ruledtabular}
				\begin{tabular}{c c c l p{17em}}
					$E$, keV & $S(E)$ & $S(0)$ & \hfil Reference & \hfil Method/project \\
					\hline
					$135$ & $51(15)$ & --- & Switkowski \textit{et al.} \cite{Switkowski1979}, 1979 & $\gamma$-ray Ge(Li) spectrometers for proton bombarding energies
					200 -- 1200 keV
					\\
					
					$340$ & $43(3)$ & --- & Ostojic \textit{et al.} \cite{Ostojic1983}, 1983 & Direct radiative capture \\
					
					$40$ & $65$ & $65$ & Cecil \textit{et al.} \cite{Cecil1992}, 1992 & Thick-target $\gamma$-ray-to-charged-particle branching ratio measurements \\
					
					$35$ & $40(14)$ & --- & Bruss \cite{Bruss1993}, 1993 & \\
					
					--- & --- & $79(18)$ & Prior \textit{et al.} \cite{Prior2004}, 2004 & Polarized proton beams, TUNL \\
					
					$250$ & $95(10)$ & --- & He \textit{et al.} \cite{He2013}, 2013 & 320 keV platform with highly charged ions \\
					
					$60$ & $92(6)$ & $95(9)$ & Piatti \textit{et al.} \cite{Piatti2020}, 2020 & LUNA collaboration \\
				\end{tabular}
			\end{ruledtabular}
		\end{table*}
	\end{widetext}
	\begin{widetext}
		\begin{table*}[h!]
			\caption[Theoretical calculations of the astrophysical $S-$factor of $^6$Li$(p, \gamma)^7$Be.]{Theoretical calculations of the astrophysical $S-$factor of $^6$Li$(p, \gamma)^7$Be.}
			\label{tab:S-factor_theory}
			\begin{ruledtabular}
				\begin{tabular}{c l l}
					$S(0)$, eV$\cdot$b	& \hfil Reference & \hfil Model \\
					\hline
					$106$ & Barker \cite{Barker1980}, 1980 & Direct-capture potential model \\
					
					$105$ & Arai \textit{et al.} \cite{Arai2002}, 2002 & Four-cluster microscopic model \\
					
					$95,5$ & Huang \textit{et al.} \cite{Huang2010}, 2010 & Single-particle model \\
					
					$114$ & Dubovichenko \textit{et al.} \cite{Dubovichenko2010,Dubovichenko2011}, 2010, 2011 & Modified potential cluster model \\
					
					$73^{+56}_{-11}$ & Xu \textit{et al.} \cite{Xu2013}, 2013 & Direct-capture potential model \\
					
					$88,34$ & Dong \textit{et al. }\cite{Dong2017}, 2017 & Gamow shell model \\
					
					$103,9$ & Gnech and Marcucci \cite{Gnech2019}, 2019 & Potential cluster model \\
					
					$96,5 \pm 5,7$ & Kiss \textit{et al.} \cite{Kiss2021}, 2021 & Modified two-body potential method  \\
					
					$92 \pm 12$ & Kiss \textit{et al.} \cite{Kiss2021}, 2021 & Modified two-body potential method  \\
					
					$98,3$ & Present work & Modified potential cluster model \\
					
				\end{tabular}
			\end{ruledtabular}
		\end{table*}
	\end{widetext}
	
\begin{figure}[h!]
		\includegraphics[width=0.5\textwidth]{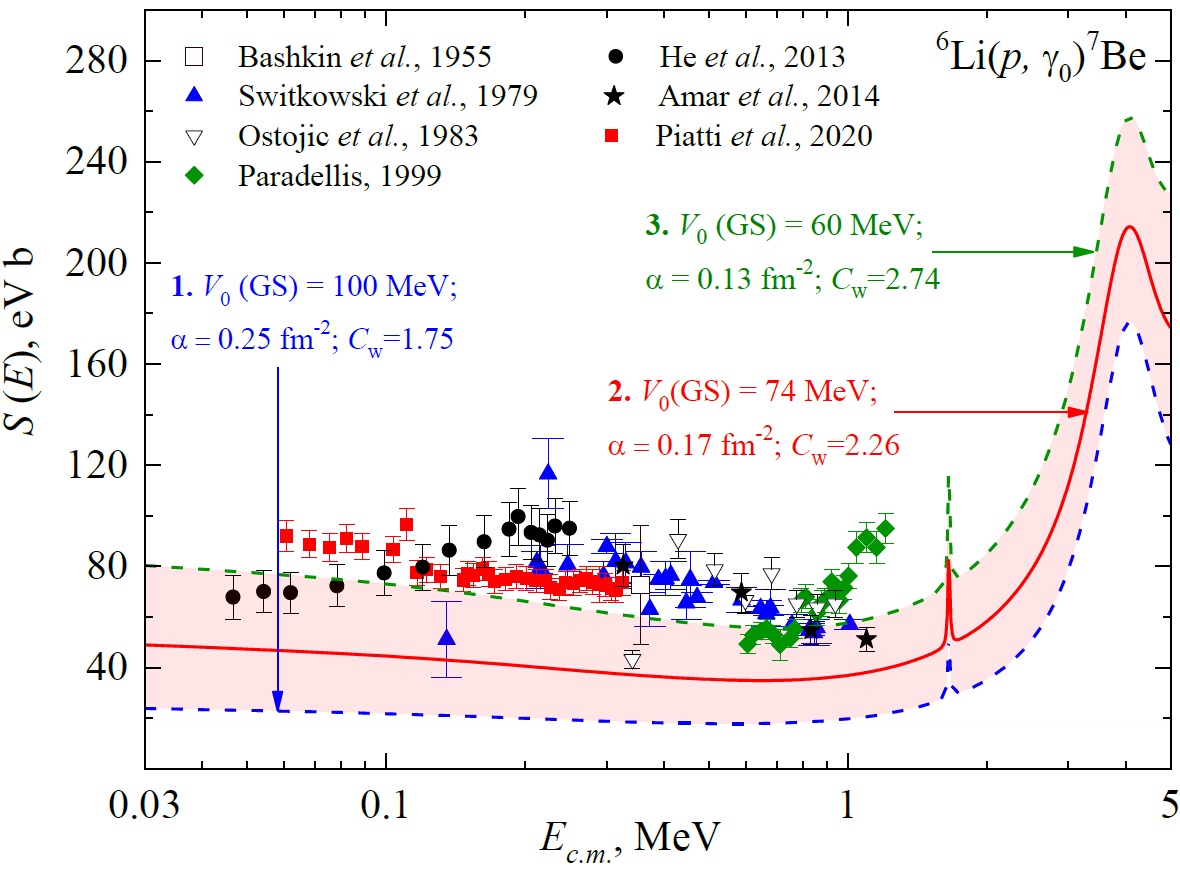}
		\caption{Astrophysical $S-$factor of the $^{6}$Li$(p,\gamma %
			_{0})^{7} $Be capture to GS. Experimental data are taken from {
				\color{Blue}$\medblacktriangleup$} -- \cite{Switkowski1979}, $
			\medblackcircle$ -- \cite{He2013}, {\color{Red}$\medblacksquare$} --
			\cite{Piatti2020}, {\color{Green}$\medblackdiamond$} -- \cite%
			{Paradellis1999}, $\medblackstar$ -- \cite{Amar2014}, $\medsquare$ -- \cite{Bashkin1955},
			$\medtriangledown$ -- \cite{Ostojic1983}. Parameters of the
			continuum potential are given in Table \ref{tab:spectrumGS}. A band
			shows the sensitivity to changes in $C_{\text{w}}$.}
		\label{fig:S-factor_GS}
	\end{figure}

\begin{figure}[h!]
		\includegraphics[width=0.5\textwidth]{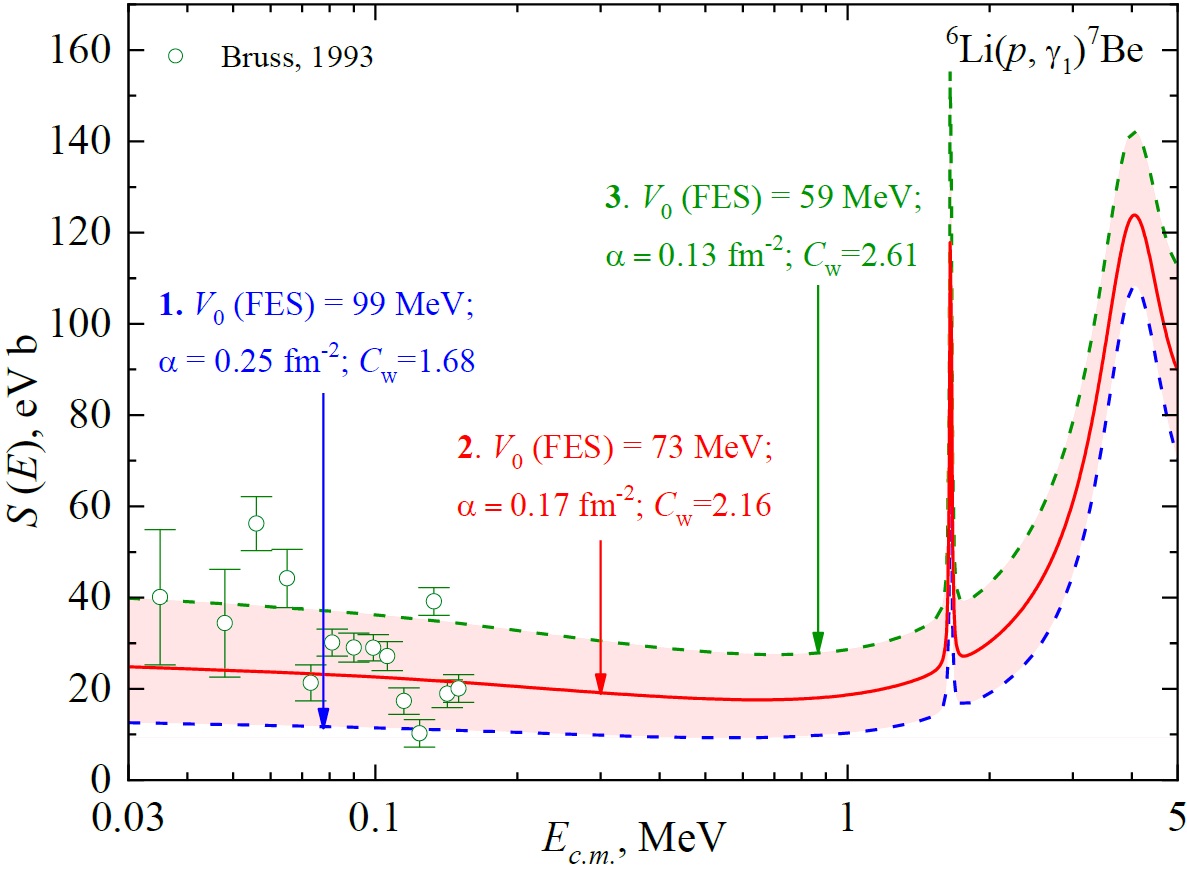}
		\caption{Astrophysical $S-$factor of the $^{6}$Li$(p,\gamma %
			_{1})^{7} $Be capture to FES. Experimental data {\color{Green}$\medcircle$}
			are taken from \cite{Bruss1993}. Parameters of the continuum
			potential are given in Table \ref{tab:spectrumFES}. A band shows the
			sensitivity to changes in $C_{\text{w}}$.}
		\label{fig:S-factor_FES}
	\end{figure}

\begin{figure}[h!]
		\includegraphics[width=0.5\textwidth]{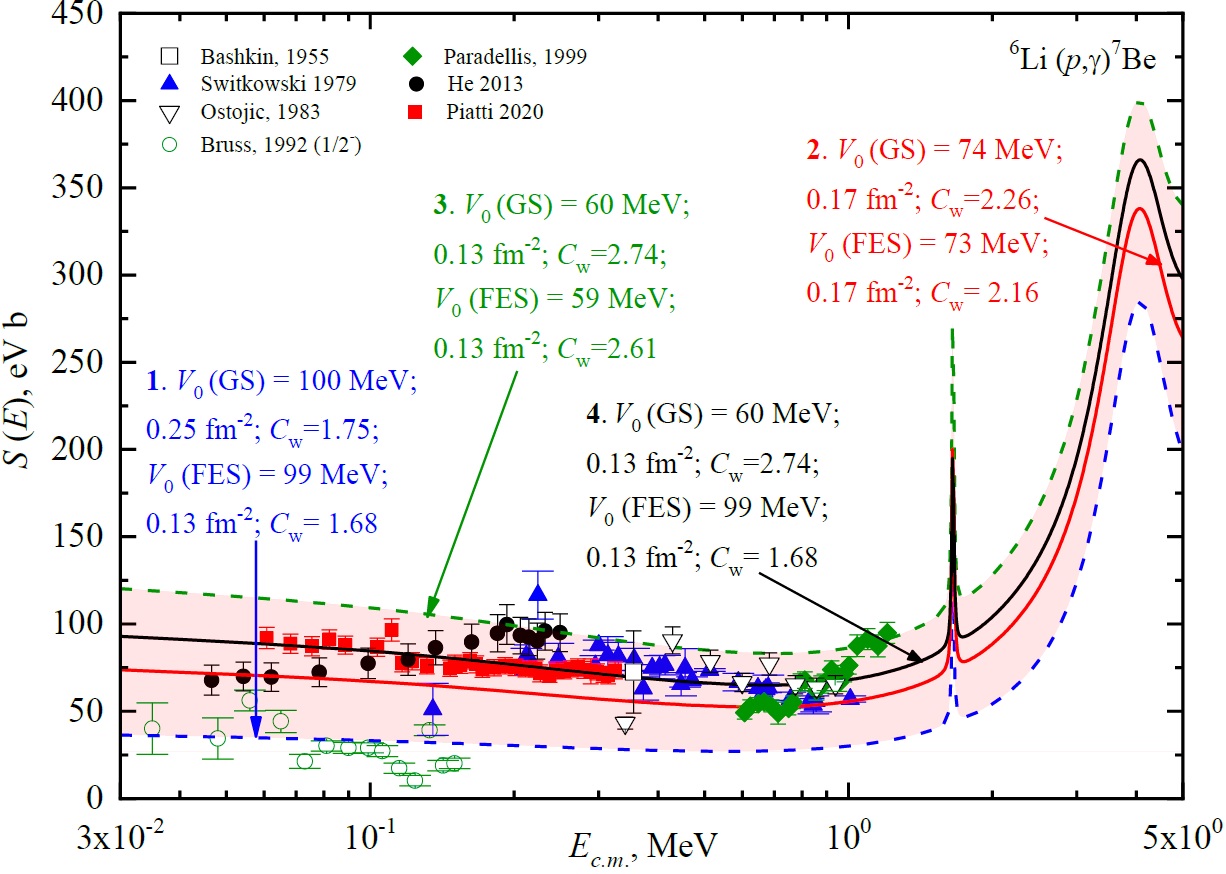}
		\caption{Astrophysical $S-$factor of the $^{6}$Li$(p,\gamma
			_{0+1})^{7}$Be capture to GS and FES. Experimental data for capture are from
			{\ \color{Blue}$\medblacktriangleup$} -- \cite{Switkowski1979}, $
			\medblackcircle$ -- \cite{He2013}, {\color{Red}$\medblacksquare$} --
			\cite{Piatti2020}, {\color{Green}$\medblackdiamond$} -- \cite
			{Paradellis1999}, $\medsquare$ -- \cite{Bashkin1955},
			$\medtriangledown$ -- \cite{Ostojic1983}, $\medblackstar$ -- \cite{Amar2014}, {\color{Green}$\medcircle$} -- \cite{Bruss1993}. The parameters of GS and FES potentials are listed in the figure. A band shows the sensitivity to changes in $C_{\text{w}}$.}
		\label{fig:S-factor_total}
	\end{figure}

	The results of the present calculations of $S-$factors along
	with available experimental data are shown in Figs. \ref{fig:S-factor_GS}, %
	\ref{fig:S-factor_FES}, and \ref{fig:S-factor_total}. Fig. \ref{fig:S-factor_GS} shows the astrophysical $S(E)-$factor of $^{6}$Li$%
	(p,\gamma _{0})^{7}$Be capture to the GS of the $^{7}$Be nucleus in an
	energy range up to 5 MeV. The solid red curve 2 and the two dashed curves,
	blue 1 and green 3, show the calculation for all transitions to the GS given
	in Table \ref{tab:spectrumGS}. Parameters $V_0$ (only integer values) and $\alpha$ of the
	corresponding potentials as well as $C_{\text{w}}$ from Table \ref{tab:Parameters} are indicated in the figures, and the parameters of the GS potentials are taken from Table \ref{tab:spectrumGS}. The solid red curve 2 is the result for the potential with the set of
	parameters \#2 from Table \ref{tab:Parameters}, leading to the average value
	of AC.
	
	In Fig. \ref{fig:S-factor_FES}, similar curves show the results
	for transitions and potentials from Table \ref{tab:spectrumFES} to FES. FES potentials have three sets of parameters from Table \ref{tab:Parameters}.
	The solid red curve 2 shows the results for capture with set \# 5 from Table
	\ref{tab:Parameters}, allowing us to determine the average value of AC.
	This result is in good agreement with the experimental data \cite{Bruss1993}
	presented in Fig. \ref{fig:S-factor_FES}. The two dashed curves 1 and 3
	almost completely cover the interval or band of cross-section errors of the
	capture to the FES.
	
	In Fig. \ref{fig:S-factor_total}, similar curves show the astrophysical $S-$%
	factor for the total cross-sections corresponding to the transition to GS
	and FES. The two dashed curves 1 and 3 show the range of $S-$factor values
	due to ambiguities in the AC of the GS and FES. For the scattering
	potentials, the parameters from Tables \ref{tab:spectrumGS} and \ref%
	{tab:spectrumFES} are used.
	
	The best agreement of the $S-$factor with experimental data is achieved for the values of $C_{\text{w}}=2.74$ for the GS and $C_{\text{w}}=1.68$ for the FES. We recommend these values as the most reliable benchmarks for future experimental studies.
	
	Fig. \ref{fig:S-factor_total} shows that almost all experimental data lie
	between the solid red curve 2 and the green dashed curve 3. If we use the GS
	and FES potentials set of parameters \#3 and \#4 from Table \ref{tab:Parameters}, the result is shown in Fig. \ref{fig:S-factor_total}
	by the black curve 4. In this case the experimental data \cite{Piatti2020} are reproduced entirely, and $S-$factor at 10 keV is found to be 101 eV$\cdot $b.
	For the scattering potentials, the data from Table \ref{tab:spectrumGS} and Table
	\ref{tab:spectrumFES} are used.
	
	Due to the uncertainty of the $S-$factor that arises from the
	uncertainty of the AC, it is desirable to select other options for the
	potentials of the GS and FES to correctly describe the LUNA data \cite%
	{Piatti2020}. This can be the subject of future work if more accurate data is
	compiled for the AC of the ground and first excited states of the $^7$Be
	nucleus.
	
	The approximation of the $S-$factor shown by the black curve 4 in Fig. \ref%
	{fig:S-factor_total} has an analytical form
	\begin{equation}
		S(E)=S_{0}+S_{1}E+S_{2}E^{2}
		\label{eq:13S-factor_analytical_form}
	\end{equation}%
	with parameters $S_{0}=98.31$ eV$\cdot $b, $S_{1}=-187.18$ MeV$^{-1}\cdot $eV%
	$\cdot $b and $S_{2}=442.51$ MeV$^{-2}\cdot $eV$\cdot $b. This approximation
	leads to $\chi ^{2}=2.4\cdot 10^{-4}$ with the error of 5\% in the energy range of 30 to 100 keV.
	This shows that $S(0)=98.3$ eV$\cdot $b and $S(30)=93.1$ eV$\cdot $b.
	
	New experimental data from LUNA \cite{Piatti2020} can be approximated to the
	first order
	\begin{equation}
		\label{eq:14Luna_approx}
		S(E)=S_0 + S_1 E
	\end{equation}
	with parameters $S_0 = 91.952$ eV$\cdot$b and $S_1 = -75.471$ MeV$^{-1}\cdot$%
	eV$\cdot$b, leading to $\chi^2 = 0.6$ and $S(0) = 92$ eV$\cdot$b.
	
	To compare the calculated $S-$factor at zero energy (10 keV), we present the
	known results for the total $S(0)$: $79(18)$ eV$\cdot $b \cite{Prior2004}, 105
	eV$\cdot $b (at 10 keV) \cite{Arai2002} and 106 eV$\cdot $b \cite{Barker1980}%
	.The $S-$factor for transitions to the ground state in \cite{Cecil1992}, 39
	eV$\cdot $b is specified, and for the transition to the first excited state,
	the $S-$factor value is equal to 26 eV$\cdot $b, the total $S-$factor is 65
	eV$\cdot $b. In our previous works \cite{Dubovichenko2010,Dubovichenko2011},
	a value of 114 eV$\cdot $b was obtained. The summary for $S-$factor
	experimental and theoretical values are presented in Tables \ref%
	{tab:S-factor_experimental} and \ref{tab:S-factor_theory}.
	
\begin{figure}[h!]
		\includegraphics[width=0.5\textwidth]{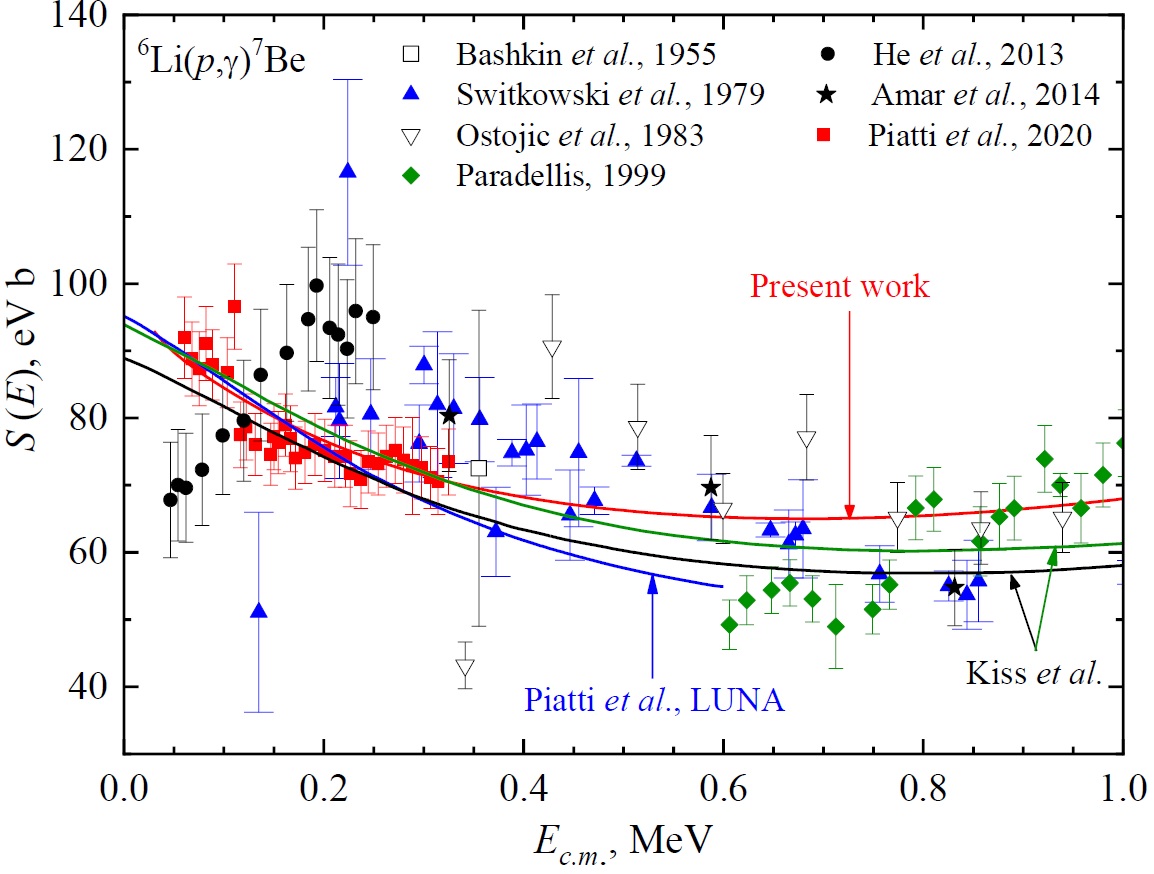}
		\caption{Comparison of $^{6}$Li$(p,\gamma )^{7}$Be reaction
			astrophysical $S-$factors. Experimental data for capture to GS and FES are
			from {\color{Blue}$\medblacktriangleup$} -- \cite{Switkowski1979}, $
			\medblackcircle$ -- \cite{He2013}, {\ \color{Red}$\medblacksquare$}
			-- \cite{Piatti2020}, {\color{Green}$\medblackdiamond$} --
			\cite{Paradellis1999}, $\medsquare$ -- \cite{Bashkin1955}, $\medtriangledown$ -- \cite{Ostojic1983}, $\medblackstar$ -- \cite{Amar2014}. Results of
			calculations: red curve -- present work; blue curve -- Ref. \cite
			{Piatti2020}; black and green curves -- Ref. \cite{Kiss2021}.}
		\label{fig:S-factor_comparison}
	\end{figure}

	To sum up, our astrophysical $S-$factor is given in Fig. \ref{fig:S-factor_comparison} with a solid red curve, together with experimental
	data and theoretical calculations. The $R-$matrix fit of the data from LUNA
	collaboration \cite{Piatti2020} and Switkowski \textit{et al.} \cite{Switkowski1979} is represented
	with the solid blue curve. A solid green curve was obtained by Kiss \textit{et al.}
	\cite{Kiss2021} using the weighted means of the ANCs from the analysis
	of the $^{6}$Li($^{3}$He,$d$)$^{7}$Be transfer reaction within the modified
	two-body potential method (MTBPM). In addition, \cite{Kiss2021} contains the
	results for the $S-$factor of the $^{6}$Li$(p,\gamma )^{7}$Be reaction
	calculated within the MTBPM, using the values of ANCs obtained from the
	analysis of the experimental astrophysical $S-$factors of the $^{6}$Li$%
	(p,\gamma )^{7}$Be reaction \cite{Piatti2020}. These results are given in
	Fig. \ref{fig:S-factor_comparison} with the solid black curve.
	
	To calculate the $^6$Li$(p, \gamma)^7$Be capture reaction rate in units of cm
	$^3$mol$^{-1}$s$^{-1}$, we used the expression
	\cite{Angulo1999} analogous to Eq. (\ref{eq:11Reaction_rate}), but substituting the corresponding constants values
	\begin{subequations}
		\begin{equation}
			\label{eq:15aReaction_rate_const}
			N_A\langle \sigma \nu \rangle = 3.7313 \cdot 10^{4} \mu^{-1/2}
			T_9^{-3/2}\int\limits_0^{\infty} \sigma (E) E \exp(-11.605 E / T_9) dE.
		\end{equation}
		In Eq. (\ref{eq:15aReaction_rate_const}) $E$ is given in MeV, the total cross-section $\sigma (E)$ is taken in $\muup$b, and $\mu$ is the reduced mass in amu and $T_9 = 10^9$ K \cite
		{Angulo1999}. Using real integration limits $E_{min}$ and $E_{max}$ Eq. (\ref{eq:15aReaction_rate_const}) becomes
		\begin{equation}
			\label{eq:15bReaction_rate_const}
			N_A\langle \sigma \nu \rangle = 3.7313 \cdot 10^{4} \mu^{-1/2} T_9^{-3/2}
			\int\limits_{E_{min}}^{E_{max}} \sigma (E) E \exp(-11.605 E / T_9) dE.
		\end{equation}
	\end{subequations}
	It is important to stress this fact as the choice of $E_{max}$ in Eq. (\ref{eq:15bReaction_rate_const}) may have a significant impact on the final result for the reaction rate. The reaction rate (\ref{eq:15bReaction_rate_const}) is calculated based on cross-sections, displayed in the form of $S-$factors (\ref{eq:10S-factor})
	in Figs. \ref{fig:S-factor_GS}, \ref{fig:S-factor_FES} and \ref{fig:S-factor_total} within the energy range $E_{min} = 1$ keV to $E_{max} =
	5$ MeV. The results of these calculations are plotted in Fig. \ref%
	{fig:reaction_rate}.
	
	\begin{figure}[h!]
		\includegraphics[width=0.5\textwidth]{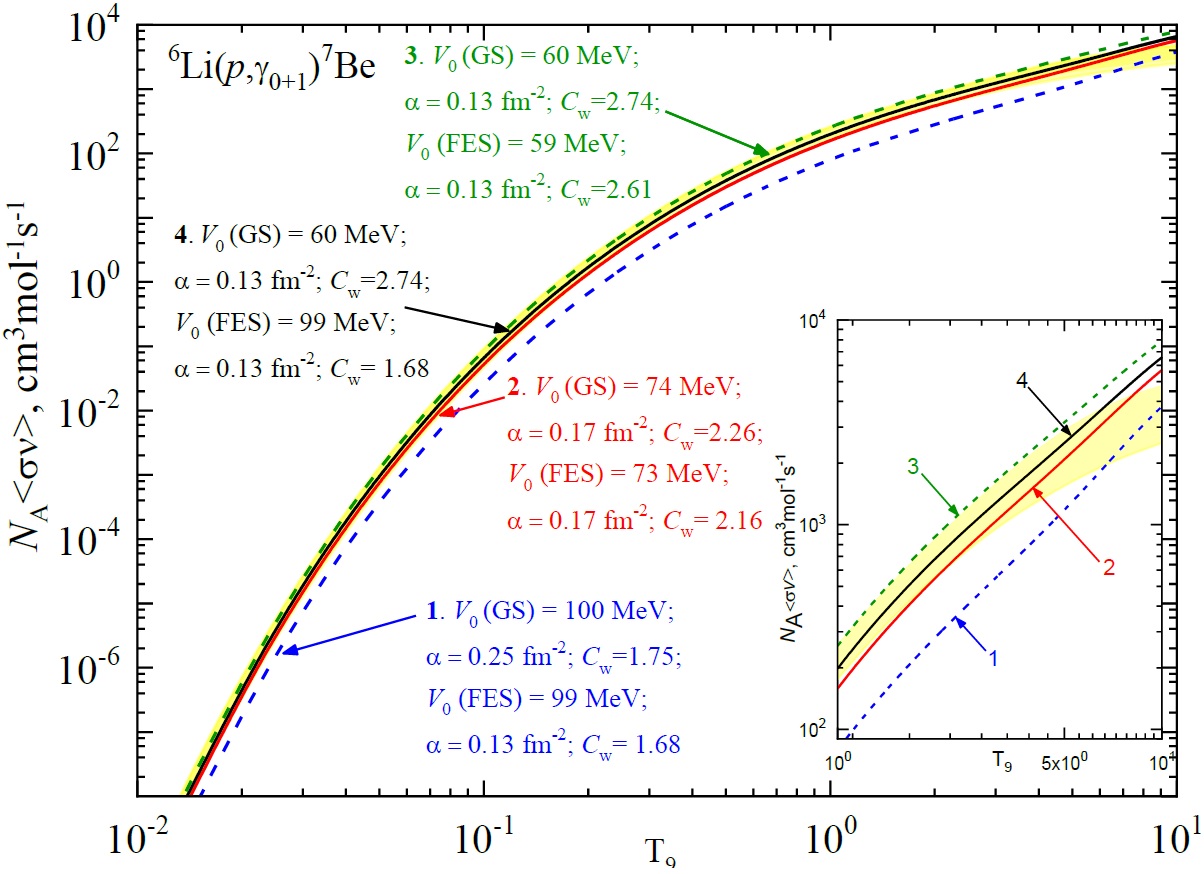}
		\caption{Total $^{6}$Li$(p,\gamma )^{7}$Be capture reaction rate.
			Curves indicate the different sums of the capture rates to the GS and
			FES. Curves are designated as in Fig. \ref{fig:S-factor_total}. The
			yellow band is taken from Ref. \cite{Xu2013}.}
		\label{fig:reaction_rate}
	\end{figure}
	
	As noted above, curve 4 in Fig. \ref{fig:S-factor_total} is in best agreement with  all experimental data of the $S-$factor. Therefore, the corresponding reaction rate, also marked as curve 4 in Fig. \ref{fig:S-factor_comparison} is most recommended description of the reaction rate. Curve 4 can be approximated by a function of the form \cite{Caughlan1988}
	
	\begin{equation}  \label{eq:16Reaction_rate_approx}
		\begin{gathered} N_A\langle \sigma \nu \rangle = a_1 / T_9^{1/3} \exp
			\left(-a_2 / T_9^{2/3} \right) \left( 1 + a_3 T_9^{1/3} + a_4 T_9^{2/3} +
			a_5 T_9 + a_6 T_9^{4/3} + a_7 T_9^{5/3} \right) + \\ + a_8 T_9^{2/3} \exp
			\left(-a_9 / T_9^{1/3}\right).
		\end{gathered}
	\end{equation}
	The parameters of approximation (\ref{eq:16Reaction_rate_approx}) with an average value of $\chi^2 = 0.014$ and the error of 5\% are given in Table \ref{tab:reaction_rate_approximation}.
	\begin{table}[h]
		\caption[The reaction rate approximation parameters.]{The reaction rate
			approximation parameters.}
		\label{tab:reaction_rate_approximation}
		\begin{ruledtabular}
			\begin{tabular}{cccccc}
				$i$ & $a_i$ & $i$ & $a_i$ & $i$ & $a_i$ \\
				\hline
				$1$ & $0.00319$ & $4$ & $-544516.9$ & $7$ & $85197.34$ \\
				$2$ & $4.16292$ & $5$ & $16401.8$ & $8$ & $924167.3$ \\
				$3$ & $3721.884$ & $6$ & $-8044.932$ & $9$ & $8.36494$ \\
			\end{tabular}
		\end{ruledtabular}
	\end{table}
	
\begin{figure}[h!]
		\includegraphics[width=0.47\textwidth]{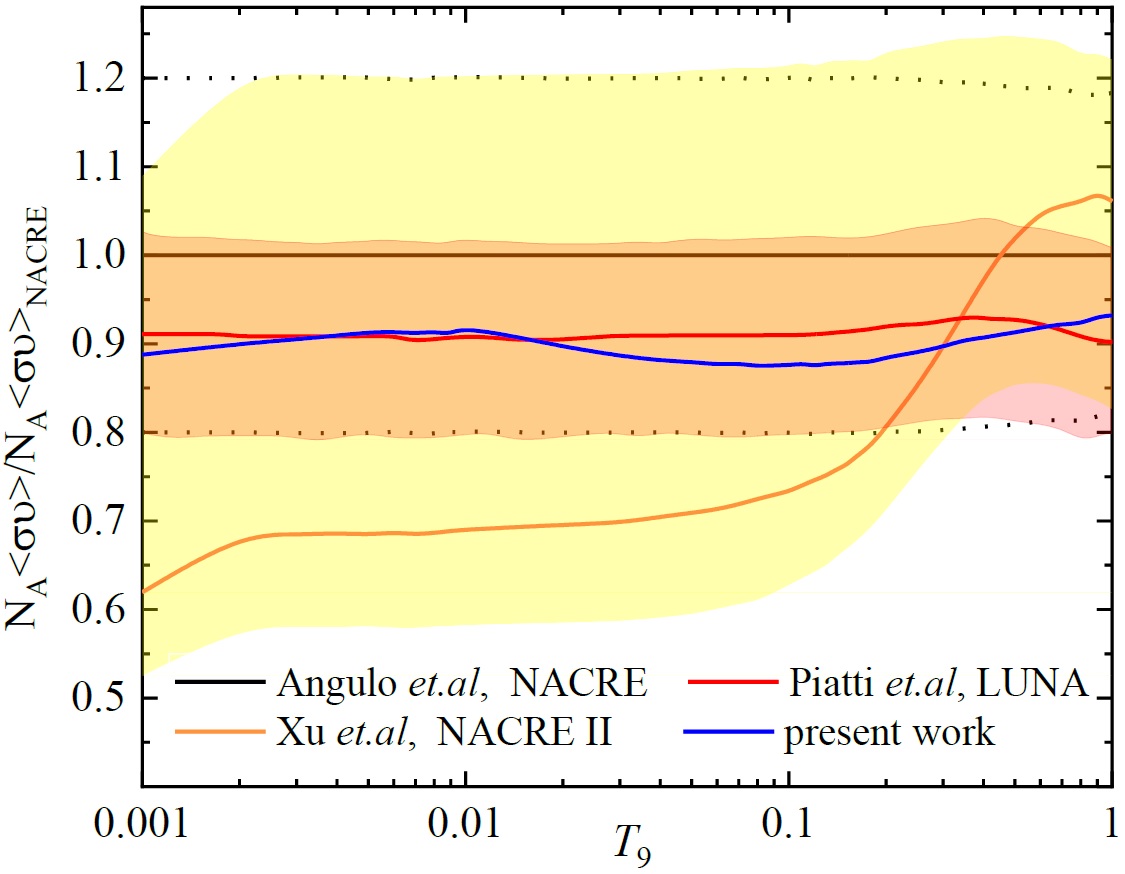}\hfill\includegraphics[width=0.47\textwidth]{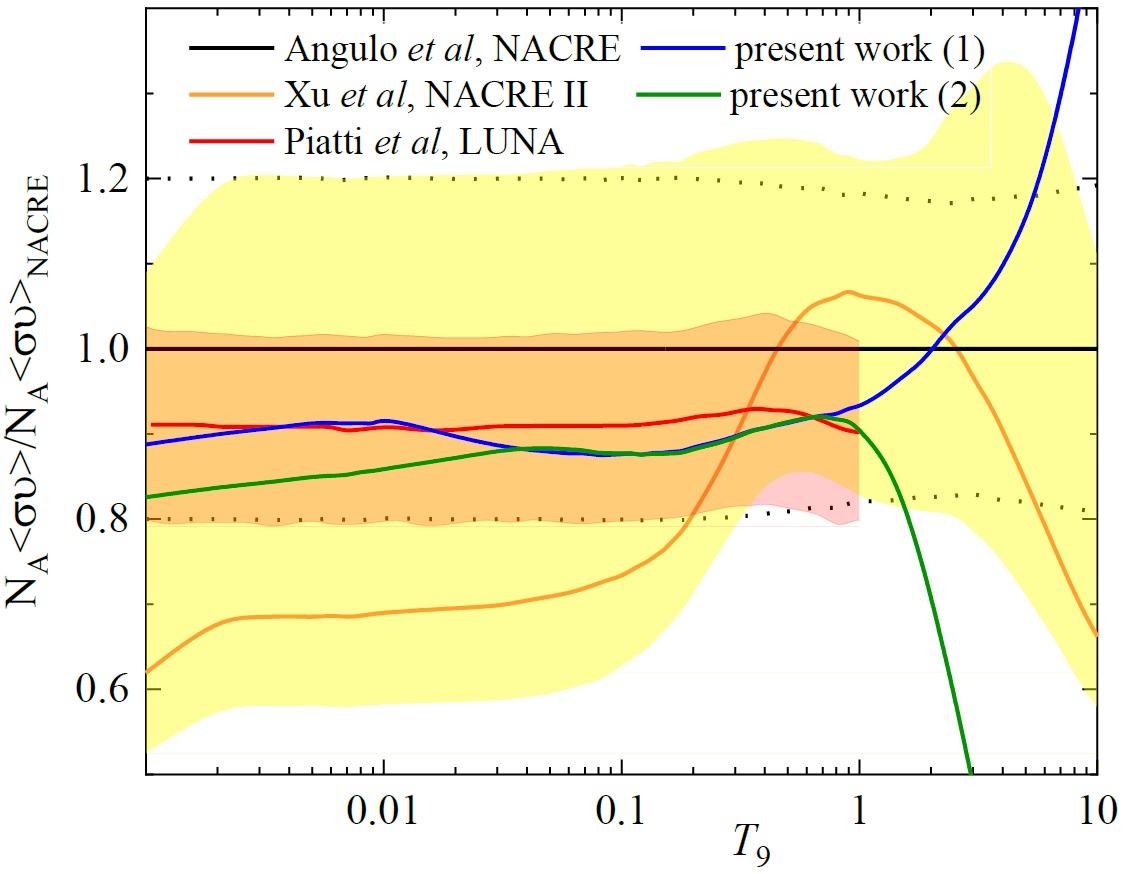}\\
		\parbox{0.47\textwidth}{\caption{Comparison of the astrophysical reaction rates in the range 0.001 to 1 $T_9$ from \cite{Angulo1999,Xu2013,Piatti2020} to present work, normalized to the NACRE rate \cite{Angulo1999}. Dotted curves represent the uncertainty of the NACRE \cite{Angulo1999} rate, while shaded areas represent the uncertainties from LUNA \cite{Piatti2020} (pale red) and NACRE II \cite{Xu2013} (yellow).}
			\label{fig:reaction_rate_comparison_1}}\hfill\parbox{0.47\textwidth}{\caption{Comparison of the astrophysical reaction rates in the range 0.001 to 10 $T_9$ from \cite{Angulo1999,Xu2013,Piatti2020} to	present work, normalized to the NACRE rate \cite{Angulo1999}. Dotted curves represent the uncertainty of the NACRE \cite{Angulo1999} rate, while shaded areas represent the uncertainties from LUNA \cite {Piatti2020} (pale red) and NACRE II \cite{Xu2013} (yellow).}
			\label{fig:reaction_rate_comparison_10}}
	\end{figure}

	Comparing our results with the reaction rate presented in NACRE \cite{Angulo1999} and NACRE II \cite{Xu2013}, we follow the format of Fig. 4 from \cite{Piatti2020}. We added our results for the reaction rate, normalized to the reaction rate from NACRE \cite{Angulo1999}. The comparison is shown in Figs. \ref{fig:reaction_rate_comparison_1} and \ref{fig:reaction_rate_comparison_10}.
	
	As stated in Ref. \cite{Piatti2020}, the LUNA "thermonuclear reaction
	rate is 9\% lower than NACRE \cite{Angulo1999} and 33\% higher than reported
	in NACRE II \cite{Xu2013} at 2 MK, and the reaction rate uncertainty has
	been significantly reduced". Fig. \ref{fig:reaction_rate_comparison_1} shows
	that the deviation between the adopted reaction rate obtained in \cite%
	{Piatti2020} and the present calculations in the range of 0.01 to 1 $T_9$
	does not exceed 5\%. Therefore, the present calculations confirm the above
	conclusion by Piatti \textit{et al.} \cite{Piatti2020}.
	
	Fig. \ref{fig:reaction_rate_comparison_10} shows two of our results --- the blue curve 1 is the reaction rate calculated on the basis of the total cross sections in the energy range from 1 keV to 5 MeV, and the green one 2 shows the rate at the upper range of the integration limit $E_{max} = 0.6$ MeV, and this value corresponded to the upper limit of energy in work \cite{Piatti2020}. The difference between these curves illustrates the importance of the resonance region contribution for the cross sections in the range of 0.6 MeV to 5 MeV at temperatures above 1 $T_9$.
	
	\section{Conclusion}
	
	\label{sec:conclusion} We present the results of calculations and analyses
	of the $S-$factor and astrophysical reaction rate for the $^{6}$Li$(p,\gamma
	)^{7}$Be reaction in the framework of MPCM. It is demonstrated that the MPCM
	approach has only one ambiguity arising from the accuracy of the
	experimentally determined asymptotic constants. This effect manifests as
	bands in Figs. \ref{fig:S-factor_GS} -- \ref{fig:S-factor_total} for the
	astrophysical $S-$factor. Precise LUNA experimental data played a role of
	the criterion, in reducing the ANC ambiguity with theoretical simulations.
	
	Comparing the $R-$matrix method which is constrained by the parameterization of the
	experimental cross-sections data, MPCM enables to implement
	calculations in wider energy ranges. We extended the energy interval for
	the total cross-sections and $S-$factors up to 5 MeV, including resonances
	in the continuum. The numerical signature of this extension is seen in Fig. \ref{fig:reaction_rate_comparison_10} for the reaction rate.
	
	It was also shown in the present work that MPCM had predicted the absence of resonance at 6.73 MeV in the nucleon channel \cite{Dubovichenko2010,Dubovichenko2011}, which was confirmed by the LUNA results \cite{Piatti2020}, as well as, indirectly, by the data from \cite{Tilley2002}.
	
	We suggest that the NACRE \cite{Angulo1999} and NACRE II \cite{Xu2013} databases should be updated in light of LUNA data \cite{Piatti2020} and present calculations.
	
	\section*{Dedication}
	We dedicate this paper to the memory of our colleague, Dr. Albert Dzhazairov-Kakhramanov, who recently passed away from COVID19.
	
	\section*{Acknowledgments}
	
	This work was supported by the grant of the Ministry of Education and
	Science of the Republic of Kazakhstan \#AP08855556 "Study of additional
	thermonuclear reactions flowing in the process of controlled thermonuclear
	fusion on lithium isotopes" through the V.G. Fesenkov Astrophysical
	Institute of the "National Center of Space Research and Technology" of
	the Aerospace committee of the Ministry of Digital Development, Innovations
	and Aerospace Industry of the Republic of Kazakhstan.
	
	\appendix
	
	\section{}
	\label{sec:appA} A solution of a two-body problem for a discrete energy
	spectrum with a given potential requires finding the binding energy of the
	system and the wave function of the state. This problem can be solved using
	the Variational Method and the Finite Difference Method (FDM) \cite
	{Dubovichenko2012}. If both methods are used for the same system of particles, it is possible to
	control the correctness of the search for the binding energy and WF of the
	state. We already have used such an approach for $p^2$H and $p^3$H systems in
	\cite{Dubovichenko2017,Dubovichenko2015b} and demonstrated that the FDM provides more precise
	description of the systems. Below we present the FDM approach.
	
	The calculation of the binding energy of a two-cluster system by the FDM
	relies on the representation of the Schr$\ddot{\text o}$dinger equation in
	finite differences \cite{Marchuk1970}. The radial equation for the central
	potential \cite{Dubovichenko2012}
	\begin{equation}  \label{eq:A1_SE_centr_potentials}
		u^{\prime \prime }_L(r)+\left[k^2 - V(r) \right] u_L(r) = 0
	\end{equation}
	with some boundary condition for $k^2 < 0$, ($k^2 = 2 \mu E / \hbar$)
	takes the form of a Sturm-Liouville type boundary value problem. Recasting the
	second derivative in finite difference form, we obtain
	\begin{equation}  \label{eq:A2}
		\begin{gathered} u'' = \left[ u_{n+1} - 2u_n + u_{n-1}\right] / h^2, \ \ \  u_n
			= u(r_n) \end{gathered}
	\end{equation}
	and (\ref{eq:A1_SE_centr_potentials}) becomes a closed system of linear
	algebraic equations. Thus, for a certain $k_0$, $D_N(k) = 0$
	\begin{equation}  \label{eq:A3}
		D_N(k) =%
		\begin{pmatrix}
			\thetaup_1 & 1 & 0 & . & . & . & 0 \\
			\alphaup_2 & \thetaup_2 & 1 & 0 & . & . & 0 \\
			0 & \alphaup_3 & \thetaup_3 & 1 & 0 & . & 0 \\
			. & . & . & . & . & . & . \\
			. & . & . & . & . & . & . \\
			0 & . & 0 & 0 & \alphaup_{N-1} & \thetaup_{N-1} & 1 \\
			0 & . & 0 & 0 & 0 & \alphaup_N & \thetaup_{N}%
		\end{pmatrix}
		=0.
	\end{equation}
	Eq. (\ref{eq:A3}) allows one to determine the binding energy $E_b$ of a system of two
	particles. The elements of the tridiagonal determinant (\ref{eq:A3}) are
	defined as follows:
	\begin{equation}  \label{eq:A4}
		\begin{aligned} &\alphaup_n = 1, & &\thetaup_n = k^2h^2 - 2 - V_n h^2, \ \ \
			n = 1, 2 , \dots, N-1, \\ &\alphaup_N = 2, & &\thetaup_N = k^2h^2 - 2 - V_N
			h^2 + 2hf(\eta,L,Z_N), \\ &Z_n = 2kr_n, & &f(k,\eta,L,Z_n) = -k -
			\dfrac{2k\eta}{Z_n} - \dfrac{2k(L-\eta)}{Z_n^2}. \end{aligned}
	\end{equation}
	Here $\eta$ is the Coulomb parameter, $k = |\sqrt{k^2}|$ is the wave number
	expressed in fm$^{-1}$ and determined by the energy of interacting particles
	in the input channel, and $V_n = V(r_n)$ is the interaction potential of
	clusters at the point $r_n = nh$ from the interval of zero to $R$.
	The number of equations $N$ or the dimension of the determinant, which
	usually turns out to be in the range $100\ 000-1\ 000\ 000$ \cite{Dubovichenko2012}, $h = \Delta r/N$ is the step of the finite difference
	grid and $\Delta r$ is the solution interval of the system (usually from
	zero to $r_N = R$).
	
	By writing $f(k,\eta,L,Z_n)$ in the form given in Eq. (\ref{eq:A4}) it is
	possible to take the Coulomb interaction into account \cite{Abramowitz1972}.
	The form of the logarithmic derivative of the WF in the external region can
	be obtained from the integral representation of the Whittaker function \cite%
	{Abramowitz1972}
	\begin{equation}  \label{eq:A5}
		f(k, \eta, L, Z) = -k - \dfrac{2k\eta}{Z} - \dfrac{2k(L-\eta)}{Z^2} S(\eta,
		L, Z),
	\end{equation}
	where
	\begin{equation}  \label{eq:A6}
		S(\eta, L, Z) = \dfrac{\int\limits_{0}^{\infty} t^{L+\eta+1} (1 +
			t/Z)^{L-\eta-1} e^{-t}dt }{\int\limits_{0}^{\infty} t^{L + \eta} (1 +
			t/Z)^{L-\eta} e^{-t}dt}.
	\end{equation}
	Calculations show that the value $S(\eta,L,Z)$ does not exceed 1.05, and its
	effect on the binding energy of a two-particle system is negligible \cite%
	{Dubovichenko2012}. When $f(k,0,0,Z) = -k$ in Eq. (\ref{eq:A5}), the binding energy search process is noticeably accelerated.
	
	The calculation of the band determinant $D_N(k)$ for a given $k$ is carried
	out using recurrent formulas of the form \cite{Marchuk1970}
	\begin{equation}  \label{eq:A7}
		\begin{aligned} &D_{-1} = 0, & &D_n = \thetaup_n D_{n-1} - \alphaup_n
			D_{n-2}, \\ &D_0 = 1, & &n = 1, \dots, N. \end{aligned}
	\end{equation}
	
	Any energy $E$ or wave number $k$ that leads to zero determinant
	\begin{equation}  \label{eq:A8}
		D_N(k_0) = 0
	\end{equation}
	is an eigenenergy of the system $E_b$ or $k_0$, and the wave function at
	this energy, determined by recurrent process below, is an eigenfunction of
	the problem.
	
	Methods for determining the zero of some functional of one variable $k$ are
	well known \cite{Korn1968}. The number $N_D$ of determinant values is
	determined automatically from the accuracy condition of the binding energy
	value. The latter one is usually set to the level $\epsilon \approx 10^{-5}$%
	--$10^{-9}$MeV, and $r_N = R$ is fixed on the range 20--30 fm \cite%
	{Dubovichenko2012}.
	
	After determining the eigenenergy $E_b$, the WF of this state is sought. To
	find the shape of the eigenfunctions of bound states, the recurrent
	procedure
	\begin{equation}  \label{eq:A9}
		\begin{aligned} &u_0 = 0, & &u_n = \thetaup_{n-1} u_{n-1} + u_{n-2}, \\ &u_1
			= \text{const}, & &n = 2, \dots, N. \end{aligned}
	\end{equation}
	is carried out, where $u_1$ is an arbitrary number, usually fixed
	on the range 0.01--0.1 \cite{Korn1968}.
	
	For bound states, the determined WF is normalized to unity. Comparing it to Whittaker asymptotics, one can find an asymptotic constant denoted by $C_%
	\text{w}$ (see Sec. \ref{sec:calculation_methods}).
	
	The WF search area $R$ is usually of 20 to 30 fm, and the number of steps
	$N_{WF}$ for the desired WF is fixed between 10 000 and 50 000. Only
	in the case of a very low binding energy (0.1--0.2 MeV) the WF search area
	increased to 100--200 fm or more.
	
	The recurrence relation (\ref{eq:A9}) is also used to search for WFs in the
	case of a continuous spectrum of eigenvalues at predetermined positive
	energy $(k^2 > 0)$ of interacting particles\cite{Dubovichenko2012}. However,
	the WF must now be matched with asymptotics of the form
	\begin{equation}  \label{eq:A10}
		N_L u_L(r) \xrightarrow[r \rightarrow R]{} F_L(kr) + \tan\left( \deltaup_{S,L}^J \right) G_L (kr).
	\end{equation}
	
	Matching the numerical solution $u_L(R)$ of Eq. (\ref{eq:A1_SE_centr_potentials}) for two points at large distances ($R$ on the
	order of 10--20 fm) with asymptotics (\ref{eq:A10}), it is possible to
	calculate the scattering phase shifts for each value of the momenta $JLS$
	for a given energy of interacting particles, as well as the normalization of
	the WF for scattering processes \cite{Dubovichenko2012}.
	To calculate the WF, one can also use the Numerov method \cite{Hairer1993}.
	When the number of steps exceeds 10 000, both methods yield the same
	results within the typical required accuracy. Such results can be compared
	by calculating the values of AC or charge radii for the BS or the matrix elements for the scattering processes \cite{Dubovichenko2012}.

	\section{}
	\label{sec:appB}
	\begin{widetext}
		\begin{table*}[h!]
			\caption[The astrophysical $^6$Li$(p, \gamma)^7$Be reaction rate in the range of 0.001 to 10 $T_9$]{The astrophysical $^6$Li$(p, \gamma)^7$Be reaction rate in the range of 0.001 to 10 $T_9$}
			\label{tab:A2_reaction_rate}
			\begin{ruledtabular}
				\begin{tabular}{cccccccc}
					$T_9$ & Rate & $T_9$ & Rate & $T_9$ & Rate & $T_9$ & Rate \\
					\hline
					0.001 & $3.20\times 10^{-29}$ & 0.035 & $6.90\times 10^{-5}$ & 0.19 & $1.35 \times 10^0$ & 2.25 & $7.80\times 10^2$ \\
					
					0.002 & $7.07\times 10^{-22}$  & 0.040 & $1.92\times 10^{-4}$ & 0.20 & $1.67 \times 10^0$ & 2.50 & $9.07\times 10^2$ \\
					
					0.003 & $2.53\times 10^{-18}$ & 0.045 & $4.57\times 10^{-4}$ & 0.25 & $3.98 \times 10^0$ & 2.75 & $1.03\times 10^3$ \\
					
					0.004 & $4.35\times 10^{-16}$ & 0.050 & $9.60\times 10^{-4}$ & 0.30 & $7.64 \times 10^0$ & 3.00 & $1.16\times 10^3$ \\
					
					0.005 & $1.68\times 10^{-14}$ & 0.055 & $1.83\times 10^{-3}$ & 0.35 & $1.28\times 10^1$ & 3.25 & $1.29\times 10^3$ \\
					
					0.006 & $2.71\times 10^{-13}$ & 0.060 & $3.25\times 10^{-3}$ & 0.40 & $1.94\times 10^1$ & 3.50 & $1.42\times 10^3$ \\
					
					0.007 & $2.49\times 10^{-12}$ & 0.065 & $5.41\times 10^{-3}$ & 0.45 & $2.75\times 10^1$ & 3.75 & $1.56\times 10^3$ \\
					
					0.008 & $1.54\times 10^{-11}$ & 0.070 & $8.55\times 10^{-3}$ & 0.50 & $3.71\times 10^1$ & 4.0 & $1.69\times 10^3$ \\
					
					0.009 & $7.20\times 10^{-11}$ & 0.075 & $1.30\times 10^{-2}$ & 0.55 & $4.81\times 10^1$ & 4.5 & $1.95\times 10^3$ \\
					
					0.010 & $2.70\times 10^{-10}$ & 0.080 & $1.89\times 10^{-2}$ & 0.60 & $6.03\times 10^1$ & 5.0 & $2.22\times 10^3$ \\
					
					0.011 & $8.58\times 10^{-10}$ & 0.085 & $2.68\times 10^{-2}$ & 0.65 & $7.37\times 10^1$ & 5.5 & $2.50\times 10^3$ \\
					
					0.012 & $2.38\times 10^{-9}$ & 0.090 & $3.69\times 10^{-2}$ & 0.70 & $8.83\times 10^1$ & 6.0 & $2.77\times 10^3$ \\
					
					0.013 & $5.92\times 10^{-9}$ & 0.095 & $4.97\times 10^{-2}$ & 0.75 & $1.04\times 10^2$ & 6.5 & $3.05\times 10^3$ \\
					
					0.014 & $1.35\times 10^{-8}$ & 0.10 & $6.55\times 10^{-2}$ & 0.80 & $1.20\times 10^2$ & 7.0 & $3.32\times 10^3$ \\
					
					0.015 & $2.83\times 10^{-8}$ & 0.11 & $1.08\times 10^{-1}$ & 0.85 & $1.38\times 10^2$ & 7.5 & $3.59\times 10^3$ \\
					
					0.016 & $5.60\times 10^{-8}$ & 0.12 & $1.67\times 10^{-1}$ & 0.90 & $1.56\times 10^2$ & 8.0 & $3.86\times 10^3$ \\
					
					0.017 & $1.05\times 10^{-7}$ & 0.13 & $2.48\times 10^{-1}$ & 0.95 & $1.74\times 10^2$ & 8.5 & $4.13\times 10^3$ \\
					
					0.018 & $1.86\times 10^{-7}$ & 0.14 & $3.52\times 10^{-1}$ & 1.00 & $1.94\times 10^2$ & 9.0 & $4.38\times 10^3$ \\
					
					0.019 & $3.18\times 10^{-7}$ & 0.15 & $4.84\times 10^{-1}$ & 1.25 & $2.99\times 10^2$ & 9.5 & $4.63\times 10^3$ \\
					
					0.020 & $5.23\times 10^{-7}$ & 0.16 & $6.48\times 10^{-1}$ & 1.50 & $4.13\times 10^2$ & 10 & $4.88\times 10^3$ \\
					
					0.025 & $4.13\times 10^{-6}$ & 0.17 & $8.45\times 10^{-1}$ & 1.75 & $5.32\times 10^2$ &  &  \\
					
					0.030 & $1.98\times 10^{-5}$ & 0.18 & $1.08\times 10^0$ & 2.00 & $6.55\times 10^2$ &  &  \\
					
				\end{tabular}
			\end{ruledtabular}
		\end{table*}
	\end{widetext}
	
	\bibliographystyle{apsrev4-2}
	
\end{document}